# Graphene oxide nanosheets disrupt lipid composition, Ca$^{2+}$ homeostasis and synaptic transmission in primary cortical neurons

*Mattia Bramini[†], Silvio Sacchetti[†], Andrea Armirotti[#], Anna Rocchi[†], Ester Vázquez[¤], Verónica León Castellanos[¤], Tiziano Bandiera[#], Fabrizia Cesca[†]\* and Fabio Benfenati[†]\**

[†]Center for Synaptic Neuroscience, Istituto Italiano di Tecnologia and Graphene Labs, Istituto Italiano di Tecnologia, 16163 Genova, Italy; [#]Drug Discovery and Development, Istituto Italiano di Tecnologia, 16163 Genova, Italy; [¤]Departamento de Química Orgánica, Universidad de Castilla La-Mancha, 13071 Ciudad Real, Spain

ABSTRACT

   Graphene has the potential to make a very significant impact on society, with important applications in the biomedical field. The possibility to engineer graphene-based medical devices at the neuronal interface is of particular interest, making it imperative to determine the biocompatibility of graphene materials with neuronal cells. Here we conducted a comprehensive analysis of the effects of chronic and acute exposure of rat primary cortical neurons to few-layers pristine graphene (GR) and monolayer graphene oxide (GO) flakes. By combining a range of cell biology, microscopy, electrophysiology and "omics" approaches we characterized the graphene-neuron interaction from the first steps of membrane contact and internalization to the long-term effects on cell viability, synaptic transmission and cell metabolism. GR/GO flakes are found in contact with the neuronal membrane, free in the cytoplasm and internalized through the endo-lysosomal pathway, with no significant impact on neuron viability. However, GO exposure selectively caused the inhibition of excitatory transmission, paralleled by a reduction in the number of excitatory synaptic contacts, and a concomitant enhancement of the inhibitory activity. This was accompanied by induction of autophagy, altered $Ca^{2+}$ dynamics and by a downregulation of some of the main players in the regulation of $Ca^{2+}$ homeostasis in both excitatory and inhibitory neurons. Our results show that, although graphene exposure does not impact on neuron viability, it does nevertheless have important effects on neuronal transmission and network functionality, thus warranting caution when planning to employ this material for neuro-biological applications.

Graphene is a carbon crystal consisting of a two-dimensional crystalline honeycomb-lattice structure made of $sp^2$-hybridized carbon atoms.[1, 2] Because of its two-dimensional shape and interesting electrical and mechanical properties[1, 3, 4] graphene is considered one of the most promising nanomaterials for biomedical applications and it is already employed in many industrial sectors such as optics, electronics and mechanics.[1, 3, 5-7] The development of large-scale synthesis procedures and the increasingly common use of graphene products lead to the critical question of nanosafety. Indeed manufactures, consumers and eventually patients are all potentially exposed to various extent to different graphene-based materials. A number of recent reports have described the toxicity of graphene and graphene-based materials in both *in vitro* and *in vivo* systems[8-13], nevertheless leaving the debate on graphene biocompatibility still open. The inconsistence among the various studies could be due to the different graphene materials used, such as monolayer or few layer graphene, graphene oxide, reduced graphene oxide and functionalized graphene.[2] Biocompatibility in fact depends not only on the *in vitro / in vivo* system adopted, but also on the physicochemical characteristics of the materials including their dimensions, shape and functionalization.[6, 14, 15] The intrinsic characteristics of graphene, especially its transparency, flexibility and high conductivity make this material particularly attractive for neuroscience applications, as it represents a promising tool for neuronal implants and bio-devices for neuro-oncology and neuro-regeneration research.[16, 17] Recently, healthy primary neurons were reported to grow and form synaptic connections on planar casted graphene-based substrates.[18] However, no data are as yet available on the potential physiological and pathological effects of the short- and long-term environmental exposure of primary neurons to pristine graphene (GR) and graphene oxide (GO) nanosheets.

On the basis of these considerations, we tested the biocompatibility of GR and GO flakes with neuronal cells. Parameters such as viability, morphology and functionality of in vitro networks of primary cortical neurons, as well as the intracellular route of internalized flakes, were first studied by employing a range of cell biology techniques, confocal microscopy, transmission and scanning electron microscopy, as well as patch-clamp electrophysiology. To understand in more detail the neurophysiological changes induced by flake exposure we investigated the protein and lipid contents of cells exposed to graphene by undertaking proteomics and metabolomics approaches. We found that both acute and chronic exposure to both GR and GO nanosheets do not affect cell viability, excitability and network formation. However, chronic GO exposure induced marked effects on neuronal network activity including a marked down-regulation of excitatory transmission accompanied by a decreased density of excitatory synaptic contacts, an opposite upregulation of inhibitory transmission and a strongly upregulated autophagic reaction. These effects were accompanied by alterations in the expression of membrane phospholipids and in a multiplicity of molecules involved in $Ca^{2+}$ homeostasis and dynamics including calcitriol/vitamin D and several $Ca^{2+}$-binding proteins that can account for the physiological effects of GO on neuronal network activity. As GO is the most largely employed graphene derivative for technological applications and the most biologically reactive, our results suggest caution in the use of this material for neurobiological applications.

RESULTS
*Synthesis and characterization of pristine graphene and graphene oxide*
The graphene-related materials used in the present study were few-layer GR and monolayer GO flakes. GR flakes were prepared by exfoliation of graphite through interaction with melamine by ball-milling treatment.[19] After exfoliation, melamine could be easily removed by filtration to obtain stable dispersions of few-layer graphene. Elemental analysis was performed in order to check any possible melamine traces into the final GR dispersion. Since 0.09 ppm melamine was found in dispersions of 0.09 mg/ml GR, melamine was included as internal control in all the experiments involving GR flakes (0.09 ppm melamine in $H_2O$). Monolayer GO was provided by the Grupo Antolin Ingeniería (Burgos, Spain) by oxidation of carbon fibers (GANF Helical-Ribbon Carbon Nanofibres, GANF®). Initial GO suspensions were washed with water to remove the presence of acids and were fully characterized. The elemental analysis gave the following results (as percent weight): 47.71 ± 0.03 wt %C, 3.04 ± 0.02 wt %H, 0.15 ± 0.01 wt %N and 0.27 ± 0.03 wt %S. Oxygen content was therefore calculated at ca. 48% by weight. Zeta-potential values were around -35 mV at physiological pH (data not shown). Dynamic light scattering (DLS) and transmission electron microscopy (TEM) analysis revealed a higher lateral size distribution of GR (500-2000 nm) compared to GO (100-1500 nm) flakes (**Figure S1a-b**). Raman spectroscopy provided a further characterization of the two materials. Graphene exhibits G and 2D modes around 1573 and 2700 $cm^{-1}$, that always satisfy the Raman selection rules while the D peak around 1345 cm-1 required a defect for its activation (**Figure S1c**). The intensity ratio between the D and the G band was calculated at different locations and used to quantify defects on the graphene surface, giving a significant low value for GR (0.22 ± 0.05; n=5) compared to GO (0.81 ± 0.05; n=5). Thermo-gravimetric analysis (TGA) was also used to quantify the degree of functionalization of graphene-related materials (**Figure S1d**). The low weight loss observed in GR (7%) corroborated the low quantity of oxygen groups generated by the exfoliation process in comparison to the values obtained for GO (46%). A detailed description of material production is available in the Methods section.

**Exposure to GR and GO nanosheets does not affect viability of primary cortical neurons**
We tested GR and GO on primary neurons isolated from embryonic brain cortices and cultured for up to 17 days *in vitro* (DIV). Cultures were incubated at 3 DIV with GR and GO at 1 and 10 µg/ml. Immunohistochemistry and fluorescence microscopy were used to evaluate GR and GO toxicity after 24 h, 96 h and 14 days of incubation to assess the effects of short- and long-term graphene exposure. Control groups were non-treated cells, cell treated with an equal volume of either 0.09 ppm melamine/$H_2O$ ("Ctrl" for GR samples) or $H_2O$ ("Ctrl" for GO samples) (**Figure 1a-d**). Propidium iodide (PI) and fluorescein diacetate (FDA) were used to detect dead and living cells, respectively. Samples were then imaged using an upright epifluorescence microscope and PI- and FDA-positive cells were quantified by imaging software analysis (**Figures S2-S4**). To evaluate network development, parallel samples were processed for confocal fluorescence microscopy using antibodies specific for the neuronal cytoskeleton (anti β-III tubulin) and Hoechst staining to visualize nuclei (**Figure 1b,d**, right panels). GR and GO flakes were visualized by reflected light imaging acquisition modality.[20] The results of this analysis clearly indicate that exposure to either GR or GO flakes did not affect cell survival or network development at any of the concentrations and exposure times tested. As a further viability test, electrophysiological analysis of resting membrane potential by patch-clamping (see below; **Table S1**), a very sensitive index of neuronal health, confirmed that the GR/GO treatments for up to 14 days do not have any detrimental effect on primary neurons.

To investigate the physical interaction of GR and GO flakes with neuronal cells, electron microscopy analysis was used. GR and GO flakes were identified by their characteristic laminated shape, high electron dense structure and typical reflected bright shadow in their proximity.[11, 21] Scanning electron microscopy (SEM) allowed us to explore the distribution of flakes in the cellular network and their interaction with the cell membrane (**Figure 1e-h**).

Cortical neurons were exposed to GR and GO flakes for 96 h and 14 days, then fixed and prepared for SEM analysis. Flakes could be seen adhering to the cell membrane (white arrowheads); however they did not seem to affect normal growth and development of the neuronal network, in accordance with the fluorescence microscopy data. It is important to note that most of the flakes in contact with the cell membrane were micro-sized flakes, suggesting that larger flakes are not taken up by cells and adopt a passive interaction behavior towards neurons.[22] Nano-flakes were instead internalized, as shown by TEM analysis (**Figure 1i-l**). A higher amount of internalized flakes was found at later time points (14 days) compared to early times of incubation (96 h); however in all cases no internal damage and/or organelles disruption was noticed. The interaction of the nanosheets with the plasma membrane occasionally led to the formation of membrane invaginations (**Figure S5**); in other cases, rare events of flakes piercing through and mechanically disrupting the cell membrane were observed (**Figure 1k**).[11, 21] As shown in **Figure 1g-j**, both GR and GO flakes (black arrowheads) were present inside cells as aggregate-like structures of various sizes and compactness. The aggregates were either freely localized in the cytosol (**Figure 1l**) or localized within membrane-bound vesicles (**Figure 1i-j**), possibly belonging to the endo-lysosomal pathway. As discussed above, the non-homogenous flake dispersion in solution may explain the formation of aggregates, even though we cannot exclude that these agglomerates are formed inside cells, upon fusion of multiple vesicles to the same intracellular structures.

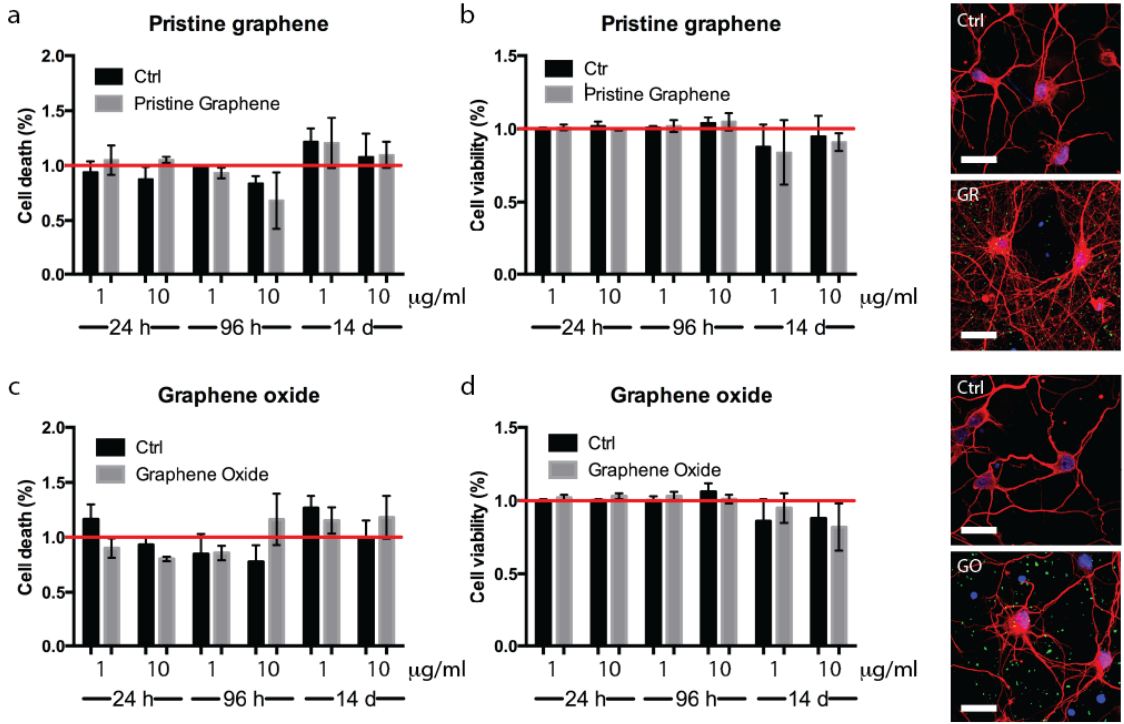
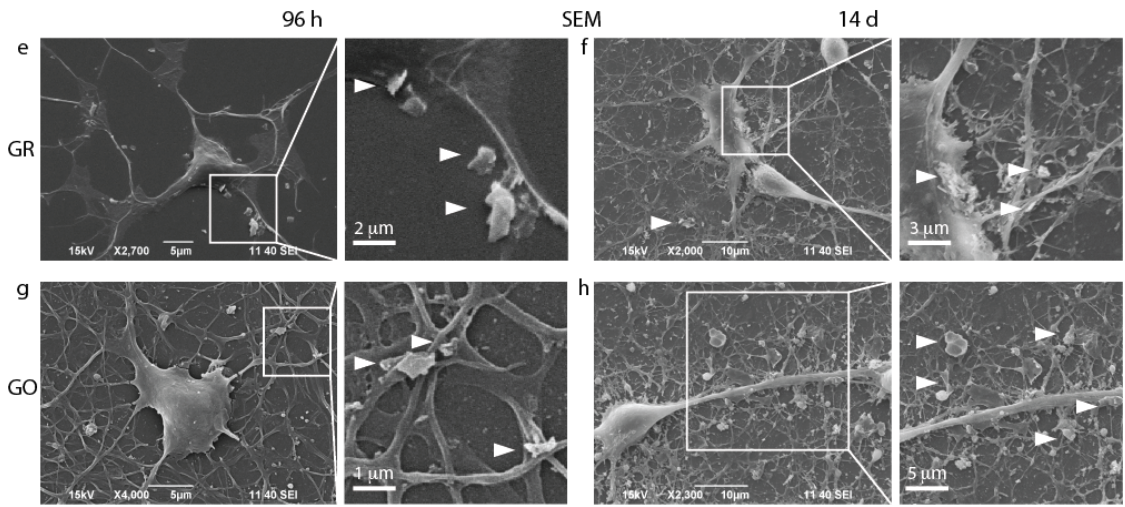
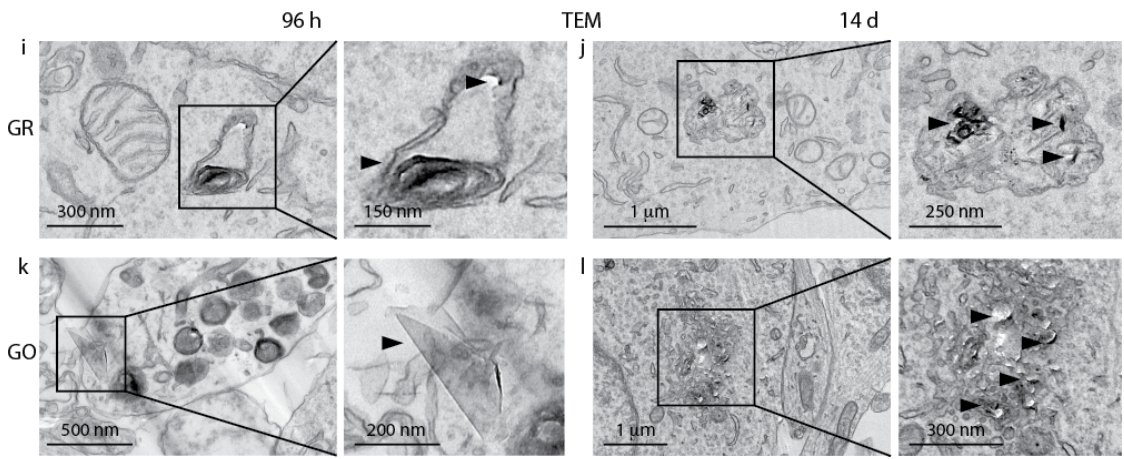

Figure 1. Cell death, cell viability and cell morphology of primary neurons exposed to GR and GO flakes. (a-d). Primary rat cortical neurons were exposed to GR (a, b) and GO (c, d) flakes (1 and 10 μg/ml) for 24 h, 96 h and 14 days, or to equivalent volumes of the respective vehicles ("Ctrl", 0.09 ppm melamine / $H_2O$ for GR flakes, $H_2O$ for GO flakes). Cell death (a, c) and viability (b, d) were evaluated by fluorescence microscopy; neurons were stained with propidium iodide (PI) for cell death quantification, fluorescein diacetate (FDA) for cell viability and Hoechst 33342 for nuclei visualization. The percentages of PI- and FDA-positive cells with respect to the total number of Hoechst-positive cells were calculated for each experimental group and normalized to the values of the untreated samples, set to 1 (red line). No statistically significant variations in cell death and cell viability were observed under all the experimental conditions (one-way ANOVA and Bonferroni's post-hoc test, 3000 cells from n=3 coverslips per experimental condition, from 3 independent neuronal preparations). In parallel samples, neurons were fixed and stained with anti-βIII tubulin antibodies (red channel) and Hoechst 33342 to visualize nuclei (blue channel), and then imaged by fluorescence confocal microscopy (b, d; right panels). GR/GO-flakes were acquired in the reflection light mode and are rendered in the green channel (scale bars, 20 μm). (e-h). SEM was used to study the interaction of flakes with neuronal cells. Neurons were exposed to GR and GO flakes for 24 h (not shown), 96 h and 14 days, fixed and prepared for SEM analysis. A large number of flakes (white arrowheads) were found in contact with the cell membrane; however, cell morphology and network development were substantially unaffected. (i-l) The cell uptake of GR and GO and the intracellular localization of graphene nanosheets were studied by TEM. At 24 h, most of the flakes were found to be outside the cells (not shown). Starting from 96 h incubation flakes were internalized into intracellular vesicles (i, j; black arrowheads), or free in the cytoplasm (l; black arrowheads).

### *GR and GO nanosheets are internalized through the endo-lysosomal pathway*

To get more insights on the identity of the intracellular vesicles containing the flakes, the internalization and intracellular fate of GR and GO flakes were investigated by confocal fluorescence microscopy (**Figure 2**). When the percentage of neuronal cells that contacted graphene nanosheets was quantified by taking into account both internalized flakes and larger aggregates in contact with the cell membrane, the large majority of neurons were found to interact with GR/GO nanosheets (mean ± SEM: GR 75.5±1.7%; GO 74.5±3.9%). We then focused on the endo-lysosomal pathway, known to be the preferential destination of nanoparticles and nanomaterials.[23-26] Neurons were fixed after 24 h, 72 h, 7 days and 14 days of GR and GO flakes exposure, and then marked with antibodies for early-endosomes (EEA1) and lysosomes (LAMP1). Three-dimensional Z-stack images were acquired by using a confocal laser-scanning microscope (CLSM). By adopting the reflection light acquisition modality[20], the visualization of GR and GO flakes was possible also in confocal mode, allowing to precisely quantify the uptake of graphene nanosheets (**Figure 2a, b, d, f, h**) and follow their intracellular location (**Figure 2c, e, g**). The evaluation of the percentages of co-localization between graphene and EEA1 or LAMP1 over the total amount of intracellular graphene revealed that GR and GO flakes are internalized with a time-dependent transition from EEA1-positive early endosomes (up to 72 h) to LAMP1-positive lysosomes, reaching a plateau value at 7 days (**Figure 2**), consistent with what found by other groups.[24-27] However, it is important to stress (**Figure 2a** and **Figure S5**) that internalization regards a relatively small percentage of total graphene materials, while the majority of GR and GO flakes are not internalized, presumably because of their relatively big size[22,24,28] and of the low phagocytic activity of neuronal cells.[29]

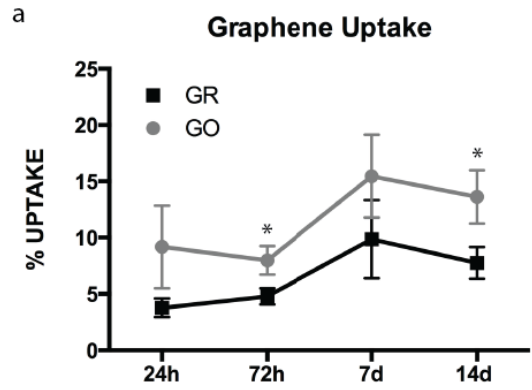
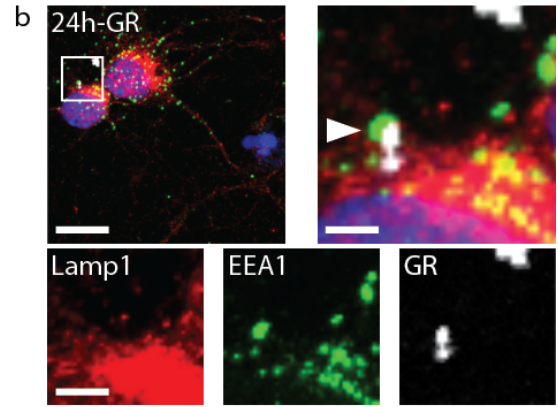
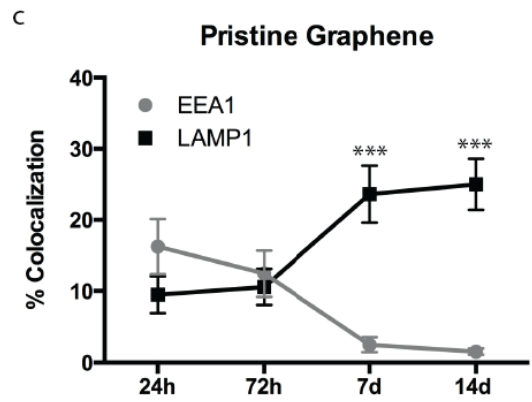
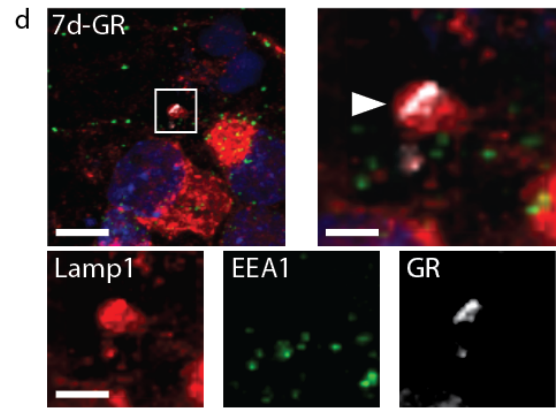
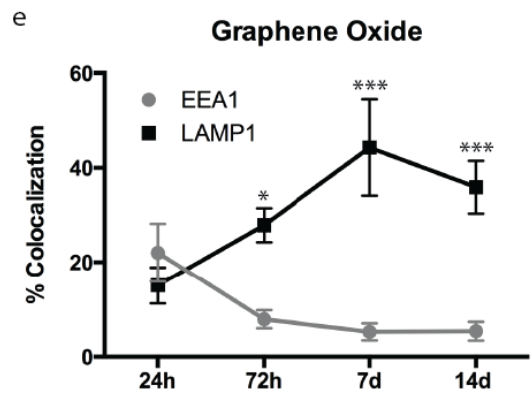
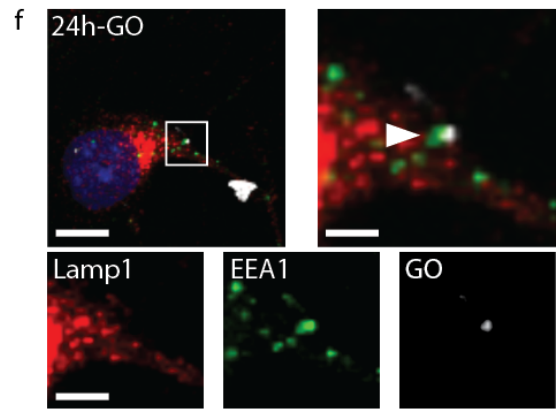
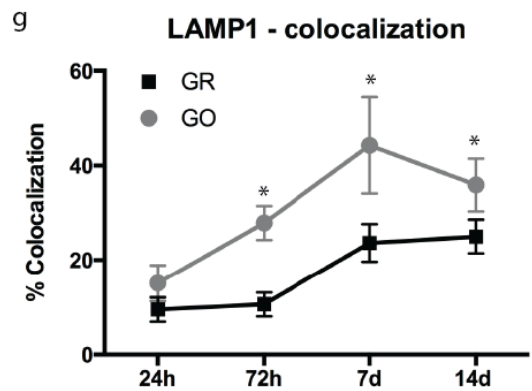
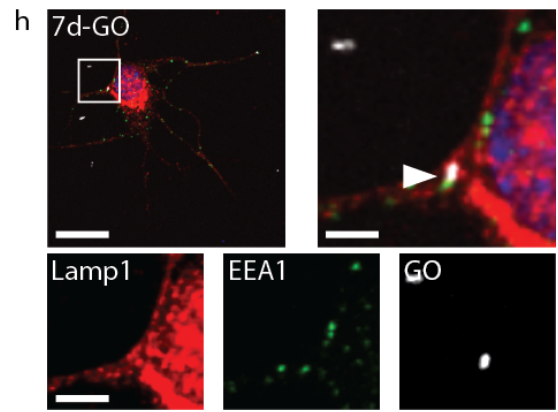

**Figure 2. Intracellular trafficking of GR and GO nanosheets.** Primary rat cortical neurons were exposed to GR and GO flakes (1 and 10 µg/ml) for 24 h, 72 h, 7 days and 14 days. Neurons were fixed and stained with anti-EEA1 and anti-LAMP1 antibodies for early endosome and lysosome identification, respectively, and with Hoechst 33342 for visualization of cell nuclei. Neurons were then imaged (**b, d, f, h**) by confocal fluorescence microscopy and three-dimensional images were acquired (scale bars, 20 µm in main images, 5 µm in zoomed images). (**a**) Total internal GR and GO flake internalization was quantified over time. (**c, e**) Quantification of the extent of co-localization of flakes with either EEA1- or LAMP1-positive organelles, analyzed using the JACOP plugin of the ImageJ software (white arrowheads; one-way ANOVA and Bonferroni's multiple comparison test, 50 cells from n=3 coverslips per experimental condition, from 3 independent neuronal preparations). (**g**) Comparison of the intracellular trafficking of GR and GO flakes shows a higher amount of GO flakes ending up in lysosomes, as a consequence of the higher internalization of GO as compared to GR flakes (**a**).

*Exposure to GR and GO nanosheets does not affect the intrinsic excitability of primary cortical neurons*

The intrinsic excitability of cortical neurons was assessed using single-cell patch-clamp recordings[30-32] to measure the cell passive and active properties in the presence of either GR or GO flakes. The resting membrane potential and the input resistance recorded in neurons exposed to 1 and 10 µg/ml of either GR or GO for 14 days were similar to untreated neurons or to vehicle-treated neurons (**Table S1**). We then assessed whether the application of various concentrations of GR/GO nanosheets impacted on the intrinsic excitability of individual primary cortical neurons by recording action potential (AP) firing in response to 500 ms current injection of stepwise increasing amplitude by whole cell patch-clamp in the current-clamp configuration (**Figure 3**). The excitability of cortical neurons was virtually unaltered by exposure to either GR or GO and the firing rate *vs* injected current response curves of GR/GO treated neurons were not significantly different from those obtained from untreated neurons or vehicle-treated neurons (**Figure 3**). The analysis of the threshold and shape of the first elicited action potential also did not show any significant treatment-dependent difference in the current or voltage threshold, AP amplitude, width, rising slope or repolarizing slope (**Table S1**). Moreover, no differences in neuronal cell functionality were observed between low (1 µg/ml) and high (10 µg/ml) GR/GO flake concentrations. These results suggest that neither graphene form affects the correct maturation and expression of the passive and active membrane properties of primary neurons.

To exclude that the lack of effect after 14 days exposure to graphene was due to graphene aggregation over time or to graphene internalization before complete cell maturation, the same experiments were performed on 14 DIV mature neurons acutely exposed to either GR or GO flakes for 24 h (**Figure S6**). Also under these conditions, no differences in the passive and active properties were found between treated and untreated neurons, confirming that graphene does not affect the basic electrophysiological properties of primary neurons, at least in the time window analyzed. We can also rule out the possibility that the preserved viability is attributable to the presence of astrocytes, as our cultures contain a very low and stable percentage of GFAP-positive cells in the in vitro period investigated (mean ± SEM: Ctrl 6.5±0.9%; GR 6.8±0.7%; GO 6.3±0.9%).

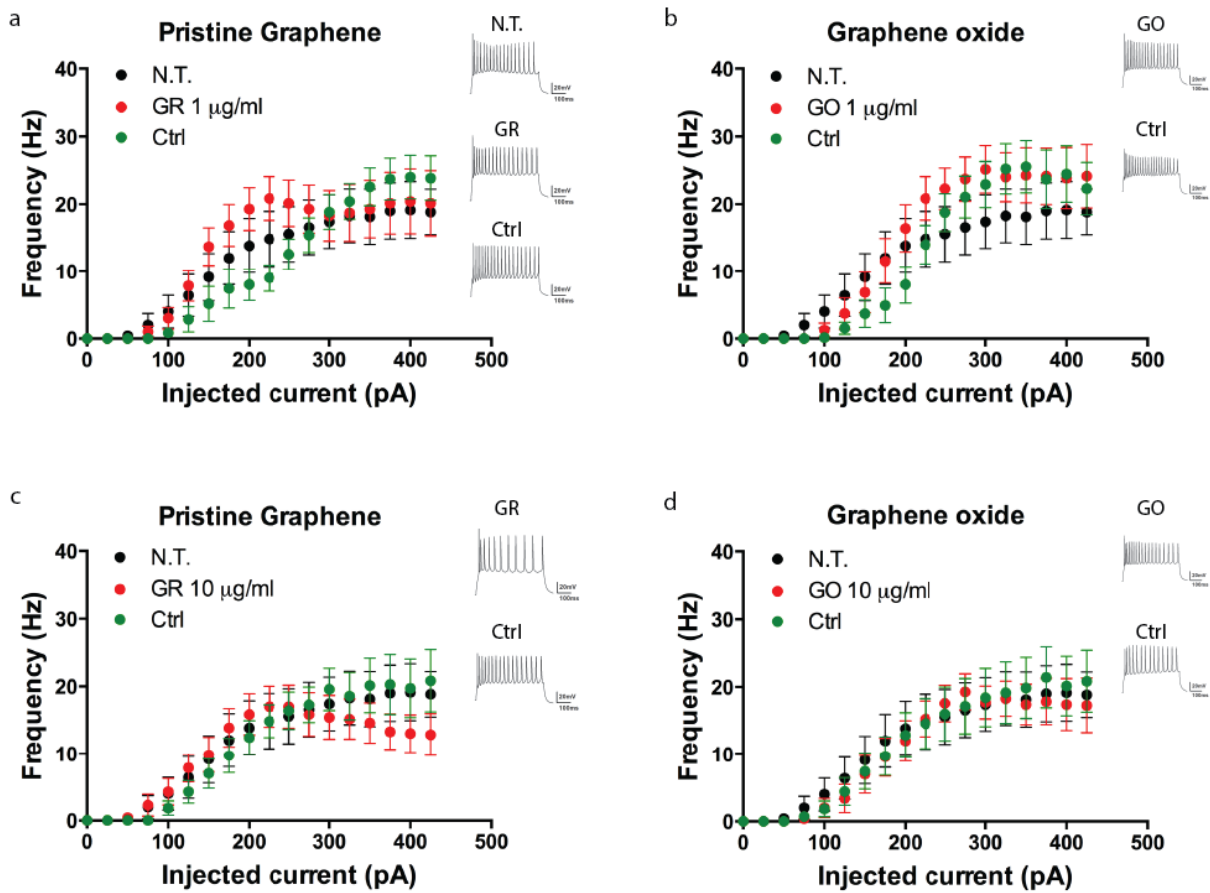

**Figure 3. Firing properties of primary cortical neurons upon chronic GR/GO flake exposure.** Neuronal cells were incubated with either GR (**a**, **c**) or GO (**b**, **d**) flakes at 1 µg/ml (**a**, **b**) and 10 µg/ml (**c**, **d**), or with the respective vehicles (Ctrl) for 14 days. Electrophysiological recordings were performed by whole cell patch-clamp in the current clamp configuration. The mean (± SEM) action potential frequency with respect to the injected current is shown. No significant differences were found between graphene-treated samples and untreated (N.T.) or vehicle-treated (Ctrl) neurons (one-way ANOVA, and Bonferroni's post-hoc test, n=20 cells per experimental condition, from 3 independent neuronal preparations). On the right of the plots, representative firing traces in response to current injections are reported.

*Exposure to GO but not GR decreases network activity and alters $Ca^{2+}$ dynamics in primary cortical neurons*

Given the substantial preservation of neuronal viability and excitability, we explored the presence of more subtle effects of graphene flakes on network activity, $Ca^{2+}$ homeostasis and synaptic transmission. We quantitatively analyzed resting $Ca^{2+}$ levels and spontaneous $Ca^{2+}$ spikes in primary cortical neurons exposed to GR/GO flakes (10 µg/ml) using live imaging with ratiometric Fura-AM. Spontaneous $Ca^{2+}$ transients reflect the firing of spontaneous action potentials by neurons present in these reverberating networks that are, in turn, the expression of the balance and interaction between excitatory and inhibitory synaptic transmission. It is also important to note that live imaging analyses predominantly investigate the activity of excitatory neurons that represent approximately 80% of the total neuronal population, as described previously.[33] Both control, GR- and GO-treated neurons displayed spontaneous Ca$^{2+}$ transients (**Figure 4a**). However, the percentage of neurons showing rhythmic activity was significantly lower in GO-treated neurons (**Figure 4b**), as were the resting Ca$^{2+}$ levels, which were decreased by GO treatment by approximately 50% with respect to vehicle-treated neurons (**Figure 4c**). Importantly, GO treatment significantly decreased the frequency of Ca$^{2+}$ spikes both under basal conditions and under conditions of hyperexcitability induced by bicuculline, a specific antagonist of GABA$_A$ receptors[34] (**Figure 4d**). Moreover, when the rise and decay of individual Ca$^{2+}$ transients were analyzed, we found that Ca$^{2+}$ transients in GO-treated neurons recovered faster than those in control neurons (**Figure 4e-g**). The same analysis, performed upon acute treatment of 14 DIV neurons with G and GO flakes for 24 h, did not reveal significant changes in any of the analyzed parameters (**Figure S7a-d**). Altogether, live Ca$^{2+}$ imaging reveals a significantly decreased network activity and complex dysregulation of Ca$^{2+}$ homeostasis.

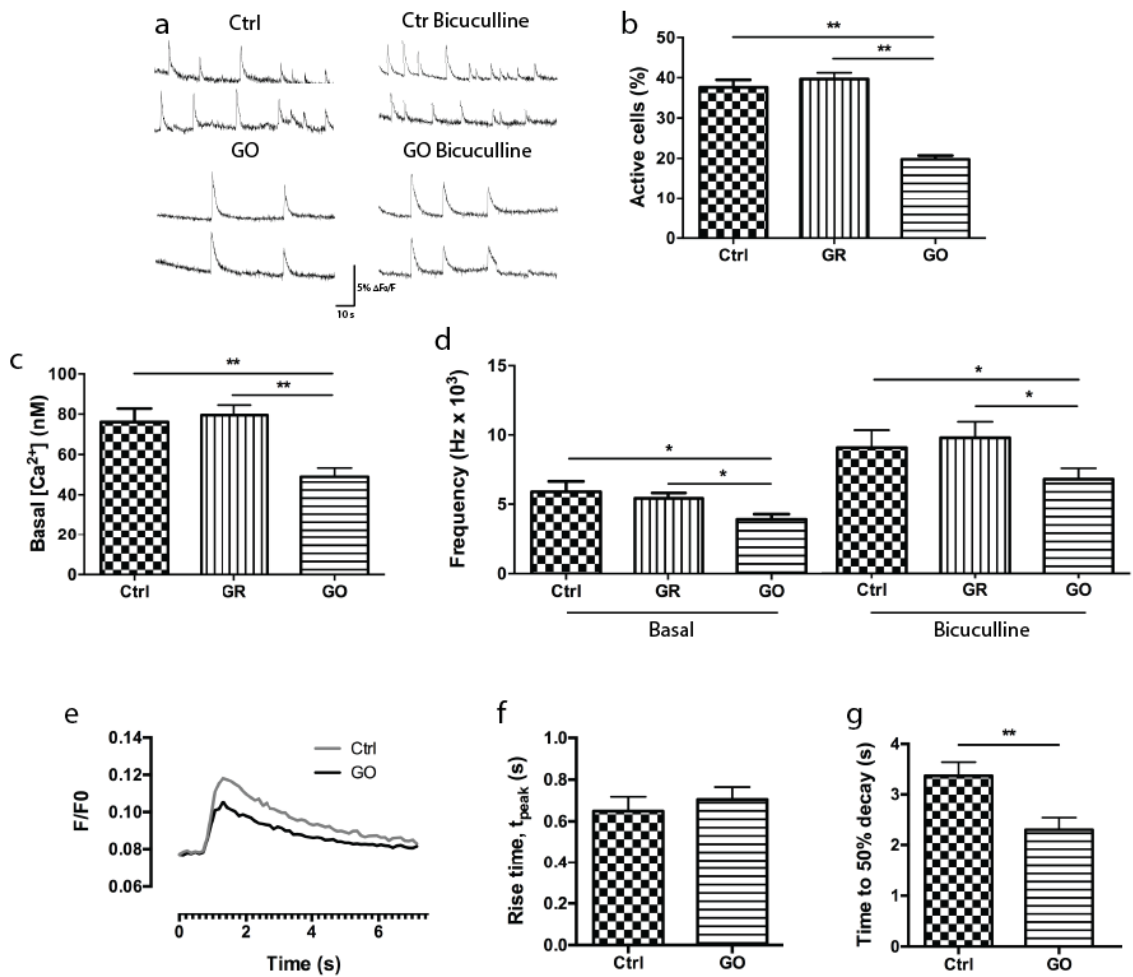

**Figure 4. Ca$^{2+}$ dynamics in primary cortical neurons treated with GO for 14 days.** (**a**) Representative spontaneous (left panels) or bicuculline-evoked (right panels) Ca$^{2+}$-oscillations recorded in 14 DIV cortical cultures treated with either vehicle (Ctrl) or 10 μg/ml GO for 14 days. (**b-c**) Percentages of spontaneously active cells (**b**) and mean resting Ca$^{2+}$ concentrations measured at the cell body level (**c**). (**d**) The frequency of Ca$^{2+}$-oscillations under basal conditions or in the presence of bicuculline (20 μM) displayed lower values for GO-treated neurons. (**e-g**) The kinetic analysis of Ca$^{2+}$ transients shows a faster recovery of Ca$^{2+}$ levels (time to 50% decay, **g**) in GO-treated neurons compared to control neurons, in the absence of changes

in the rise time (**f**). Data are means ± SEM. *p<0.05, two-tailed Student's *t*-test; n = 30 cells per experimental condition, from 4 independent neuronal preparations.

### *Exposure to GO but not GR induces an imbalance between excitatory and inhibitory synaptic transmission*

The interactions of GR/GO nanosheets with the plasma membrane, together with the low degree of flake endocytosis observed upon graphene exposure, can potentially impact on the formation, maintenance and function of synaptic connections. Synapses are in fact the sites in which the highest intensity of membrane trafficking, membrane fusion/fission events and signal transduction cascades occur in the nervous tissue.[35] Thus, we investigated whether the changes observed after treatment with GO were associated with alterations in the density of excitatory and inhibitory synapses and in the frequency and amplitude of miniature excitatory (mEPSCs) and inhibitory (mIPSCs) postsynaptic currents.

In GO-treated neuronal networks the frequency of mEPSCs was strongly depressed (approximately 75% decrease) as compared to untreated or vehicle-treated neurons (**Figure 5a, b**). Moreover, the mEPSC amplitude was also significantly decreased by approximately 50% (**Figure 5a, c**) in the absence of effects on the current rise time and decay (**Figure 5d**). Strikingly, an opposite effect was found on mIPSCs, whose frequency was significantly increased in comparison with control samples (**Figure 5e, f**), in the absence of changes in the amplitude (**Figure 5e, g**) and in the current rise time and decay (**Figure 5g**). Interestingly, these effects on excitatory and inhibitory transmission were virtually specific for GO and totally absent in neurons exposed to GR flakes (**Figure S8**). In addition, a long-term exposure to GO flakes was required to provoke the synaptic effects, since mature neurons exposed for 24 h to either GO or GR flakes did not display any synaptic phenotype (**Figure S7e-h**).

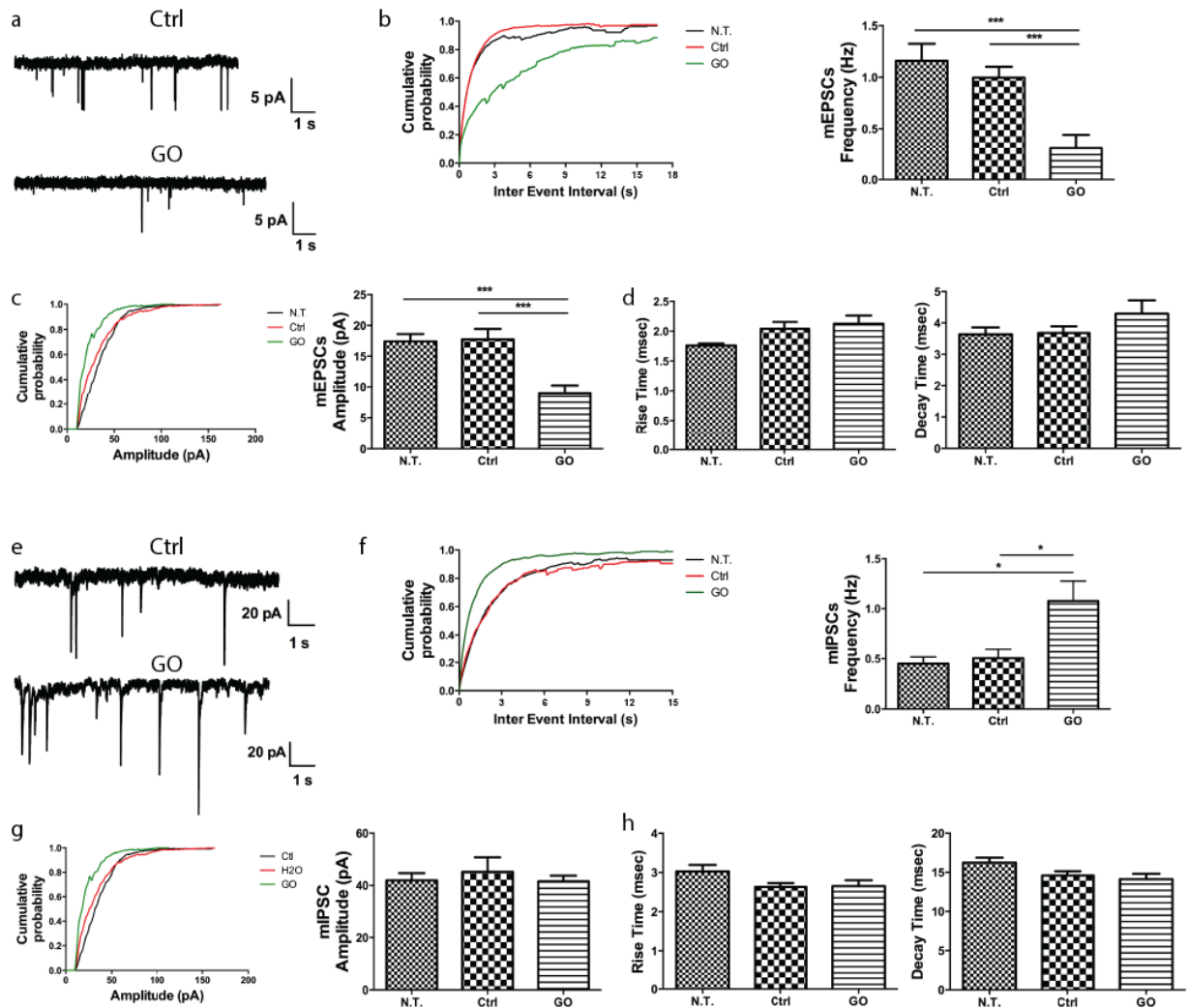

Figure 5. Activity of excitatory and inhibitory synapses in primary cortical neurons treated with GO for 14 days. (**a**) Representative recordings of miniature excitatory postsynaptic currents (mEPSCs) in cortical neurons treated with GO (10 μg/mL) or vehicle. (**b**) Cumulative distribution of inter-event intervals and mean (± SEM) frequency of mEPSCs in control and GO-treated neurons. (**c**) Cumulative distribution and mean (± SEM) of mEPSC amplitude in control and GO-treated neurons. (**d**) Mean (± SEM) values of the rise and decay times of mEPSCs (one-way ANOVA and Bonferroni's multiple comparison test; ***p<0.001; n = 15 cells from 2 independent neuronal preparations). (**e**) Representative recordings of miniature inhibitory postsynaptic currents (mIPSCs) in cortical neurons treated with GO (10 μg/mL) or vehicle. (**f**) Cumulative distribution of inter-event intervals and mean (± SEM) frequency of mIPSCs in control and GO-treated neurons. (**g**) Cumulative distribution and mean (± SEM) of mIPSC amplitude in control and GO-treated neurons. (**h**) Mean (± SEM) values of the rise and decay times of mIPSCs (one-way ANOVA and Bonferroni's multiple comparison test; *p<0.05; n = 10 cells from 2 independent neuronal preparations).

While the amplitude of miniature currents depends on the quantal size (i.e., the number of neurotransmitter molecules released by each synaptic vesicle) and on the number and sensitivity of postsynaptic receptors responsible for the current generation, the frequency of miniature currents depends on the spontaneous release probability (that is sensitive to the resting $Ca^{2+}$ levels) and on the number of synaptic contacts impinging on the patched neuron. To elucidate the latter point, we double-stained cortical cultures with antibodies for β-III tubulin to identify neuronal processes and either the vesicular glutamate transporter VGLUT1 or the vesicular GABA transporter VGAT to visualize synaptic contacts (**Figure 6**).[36-38] Indeed, treatment with GO induced a significant decrease in the density of excitatory glutamatergic contacts (**Figure 6c**), while the density of inhibitory GABAergic synapses was fully preserved (even with a slight trend toward an increase; **Figure 6d**). The impaired excitatory/inhibitory synapse ratio and the imbalance between excitatory and inhibitory transmission observed in GO-treated neurons is in line with the reduced network activity of excitatory neurons observed by live $Ca^{2+}$ imaging.

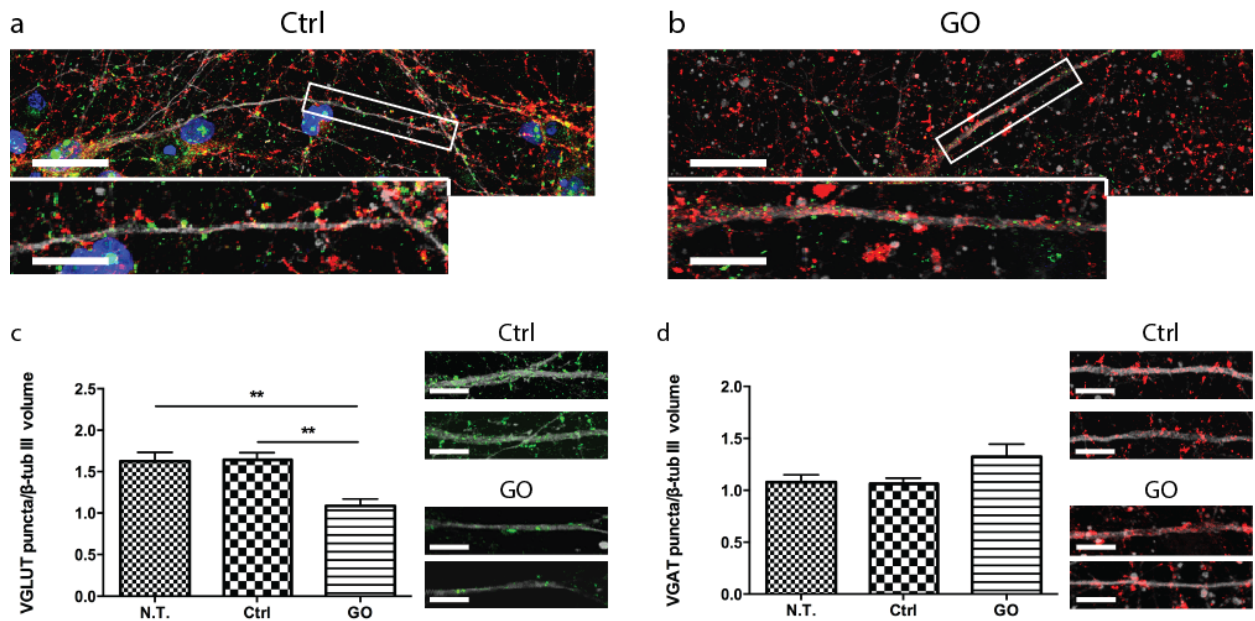

Figure 6. Synaptic density in primary cortical neurons treated with GO for 14 days. (a, b) Representative images of primary cortical neurons at 17 DIV triple stained for β-III tubulin (grey), VGLUT1 (green) and VGAT (red) are shown. Nuclei were stained with DAPI. (c, d) Density of VGLUT1-positive excitatory synapses (c) and of VGAT-positive inhibitory synapses (d) in untreated, vehicle-treated (Ctrl) and GO-treated neurons (scale bars 5 μm; one-way ANOVA and Bonferroni's multiple comparison test; *p<0.05; **p<0.01; n=30 cells per experimental condition, from 3 independent neuronal preparations).

*Exposure to GO alters the lipid and protein content of primary cortical neurons*

The extensive $Ca^{2+}$ imaging and synaptic transmission analysis revealed dramatic changes in $Ca^{2+}$ homeostasis, in conjunction with a compromised synaptic transmission upon GO exposure. To specifically investigate the effect of this specific type of graphene and get a more comprehensive understanding of the intracellular pathways that are affected by GO, metabolomics and proteomics studies were carried out on two independent preparations of primary neurons chronically treated with GO.

An untargeted high-resolution LC-MS/MS lipidomic approach allowed us to observe substantial changes in the lipid composition of neurons treated with GO flakes compared to vehicle-treated cells, including both membrane and intracellular lipids. We first used an unsupervised Principal Component Analysis[39] (**Figure 7a**) to generate a general overview of the changes induced by exposure of neurons to GO with respect to vehicle-treated neurons and GO alone (in cell culture medium). The plot describing the observed differences between the experimental groups shows a non-overlapping distribution of the three experimental groups with very limited dispersion. Using a Discriminant Analysis[39], we identified a series of phosphatidylethanolamines (PEs), which are among the major cell membrane components, that were markedly upregulated in the GO group (**Figure 7b**). Other phospholipids, such as phosphatidylinositols (PIs) and phosphatidylserines (PSs) were also differentially regulated by the exposure to GO (**Figure S9a**). Interestingly, a dramatic up-regulation of the major vitamin D3 metabolite calcitriol was also observed. These appear to be striking alterations, especially when compared to other very abundant structural lipid species, such as cholesterol, which did not vary amongst samples (**Figure 7b**). Although the total cholesterol was not affected by GO, this lipid is known to redistribute in the cell membrane in response to a variety of stimuli [40-42]. However, when neurons were incubated with the cholesterol-specific fluorescent marker filipin, no detectable changes in membrane cholesterol distribution were observed between GO-treated and untreated cells (**Figure S9d**).

We next performed proteomics on the same sample sets by digesting neuronal proteins and labeling the GO and vehicle-treated groups with TMT isobaric tags.[43] As summarized in **Figure 7c**, roughly 650 proteins were unambiguously identified in the two biological replicates, and more than 70% of them gave useful over/under expression data. Following data filtering,[44] a final list of 405 proteins was used for expression analysis (**Figure 7d** and Supporting Information Excel file). From this dataset, 16 and 57 proteins showed respectively a marked (>20% in module) up-regulation and down-regulation in the GO group and were subjected to gene ontology. Quite interestingly, calmodulin (*Calm1*) and its closest functional interactors such as $Ca^{2+}$/calmodulin-dependent kinase type II (CaMKII or *Camk2d*) and the catalytic subunit of the protein phosphatase 2B aka calcineurin (*Ppp3c*) and related proteins (such as γ-actin, *Actg2*) (**Figure 7e**) were amongst the major downregulated proteins, while a number of proteins involved in membrane trafficking and autophagy were substantially upregulated (**Tables S2 and S3**).

To validate these findings with an independent assay, we verified the up/down regulation of some proteins of interest by either qRT-PCR or western blot analysis (**Figure 7f-g**). Our data confirmed a statistically significant downregulation of a number of proteins involved in intracellular $Ca^{2+}$-dependent activities including calmodulin (*Calm1*), CaMKII and calretinin (*Calr*) and of one mitochondrial import receptor (Tom70), as well as the upregulation of proteins controlling intracellular trafficking (*T-complex protein 1 subunit delta*, TCPD; *ADP-ribosylation factor 4*, Arf4; *Nck-associated protein 1*, NCKP1/Nap125) and autophagy (*microtubule-associated proteins 1A/1B light chain 3A*, MLP3A/LC3). Notably, in the case of LC3, the active PE conjugated form (LC3II) was significantly upregulated, while its inactive precursor LC3I resulted downregulated.

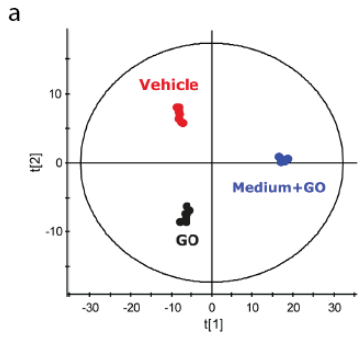
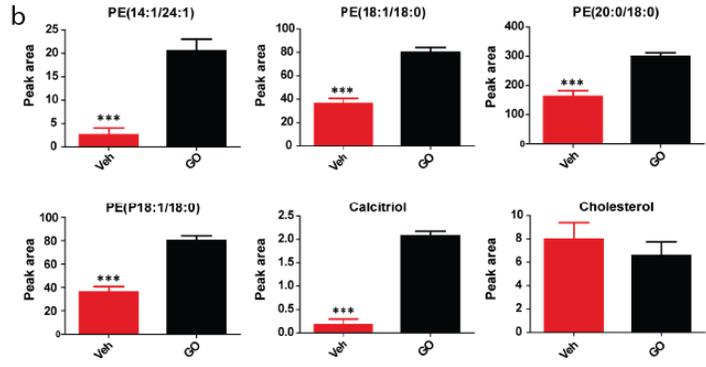
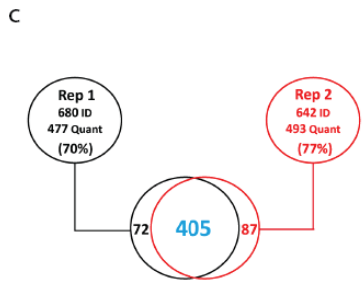
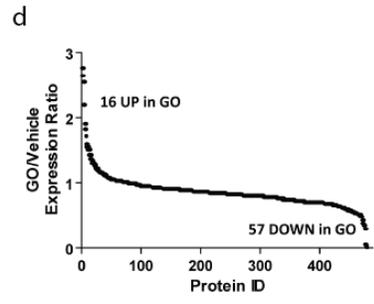
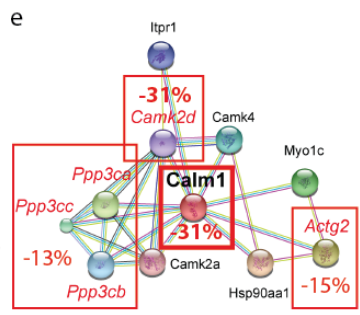
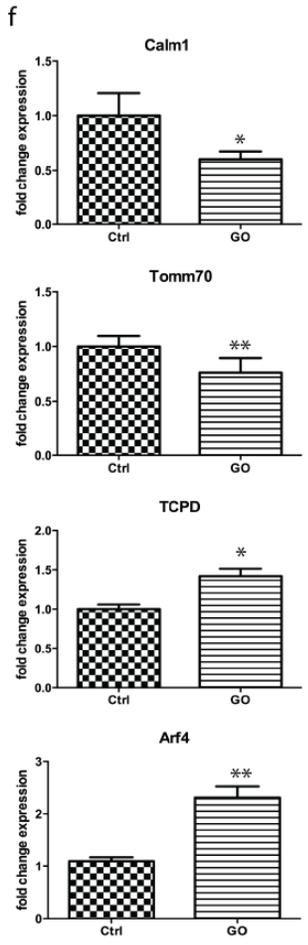
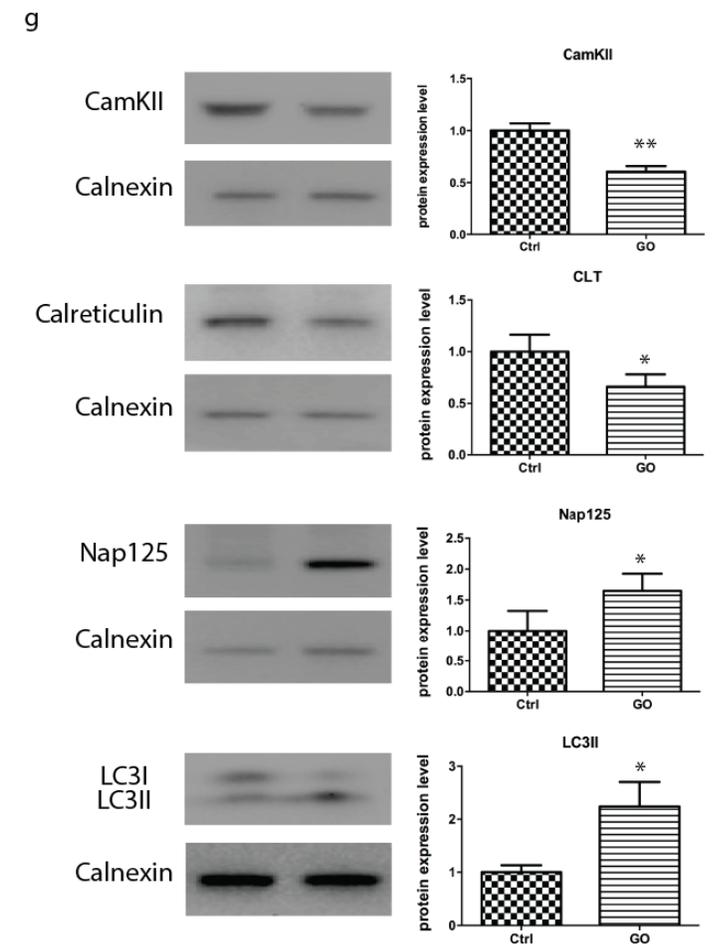

Figure 7. Lipidomic and proteomic studies on primary neurons exposed to GO for 14 days. (a) Principal component analysis[39] of the untargeted lipidomic experiment: dots representing individual neuronal samples cluster into the three experimental groups, indicating clear changes in lipid content induced by exposure to GO. Blue dots are reference samples, consisting of GO flakes in cell culture medium in the absence of cells. (b) Phosphatidylethanolamines (PEs) and calcitriol show a marked upregulation following exposure to GO, especially when compared to other invariant cell lipids, such as free cholesterol (***$p<0.001$ two-tailed Student's *t*-test). The analyte peak areas were used for quantification. These lipids are the major net contributors to the GO *vs* vehicle differences observed in (a). Values on the Y-axis are means (± SEM) normalized peak areas. (c) Results of the untargeted proteomics experiment: among the total positive protein hits, the two independent biological replicates gave ~70% of useful expression information that resulted in 405 quantified proteins. (d) Expression plot of the 405 protein hits: 16 proteins were overexpressed in GO by more than 20%, while 57 were downregulated by more than 20%. (e) Rat calmodulin (*Calm1*) network of functional interactors. The proteins reported in italic red were represented in the untargeted dataset as markedly downregulated in neurons exposed to GO (-10 to -30% in both replicates, values for each protein are reported in the red squares), clearly indicating a significant downregulation of the whole pathway. (f) Calm1, Tomm70, Tcpd and Arf4 mRNA levels were quantified by qRT-PCR in neurons treated with either vehicle (Ctrl) or GO. GUSB and HPRT1 were used as control housekeeping genes. (*$p<0.05$, **$p<0.01$, two-tailed Student's *t*-test; n = 3). (g) Western blot analysis of CaMKII, calreticulin (CLT), NCKP1/Nap125 (Nap125) and MLP3A/LC3 I/II in neurons following exposure to GO flakes. Calnexin was used as loading control. A representative experiment (left panel) and quantification (right panel; means ± SEM) are shown (*$p<0.05$, **$p<0.01$, two-tailed Student's *t*-test; n = 3).

Given the expression changes in $Ca^{2+}$-binding proteins, $Ca^{2+}$ buffers and $Ca^{2+}$-dependent enzymes, and the opposite effects observed in excitatory and inhibitory transmission, we investigated the expression levels of $Ca^{2+}$ buffer proteins that are specific for inhibitory neurons, namely parvalbumin (PV), calbindin$_{D28k}$ (CB) and calretinin (CR) by quantitative western blotting of primary neurons exposed for 14 days to either GO or vehicle. Interestingly, all three interneuron-specific $Ca^{2+}$ buffers were significantly downregulated, with a dramatic effect (4-fold reduction) for PV that is the specific $Ca^{2+}$ buffer of fast-spiking interneurons (**Figure S10**). Altogether, these data confirm the presence of a dysregulation of $Ca^{2+}$ homeostasis induced by treatment with GO that differentially affects excitatory and inhibitory neurons.

*Exposure to GO triggers macroautophagy in primary cortical neurons*
As post-mitotic cells, neurons are much more sensitive to accumulation of toxic components than replicating cells. Autophagy, macroautophagy in particular, is the only cellular degradation mechanism capable of eliminating large material aggregates, as well as expired, engulfed or damaged organelles. In the autophagic pathway, cytosolic components are sequestered in autophagosomes and degraded upon fusion with lysosomal components.[45, 46] A tight control for degradation of cytoplasmic components by the autophagy is fundamental for neuronal survival and activity under physiological conditions and inhibition of autophagy is known to cause neurodegeneration in mature neurons.[47]
An indication that chronic GO exposure could stimulate the autophagy pathway in primary neurons came from the proteomic screen that identified, among the most upregulated proteins, MLP3A/LC3, the mammalian orthologue of ATG8, which is mechanistically involved in the formation of autophagosomal vacuoles. LC3I is generated by proteolytic cleavage of pro-LC and, upon conjugation with PE, is converted into the active LC3II form that stimulates the formation of autophagosomes by specifically associating with their membranes.[46] Western blot analysis further demonstrated a specific increase in LC3II at the expense of LC3I, consistent with an activation of the autophagic pathway in neurons chronically exposed to GO (**Figure 8a**;

see also **Figure 7g**). Interestingly, activation of LC3 involves PE that was also markedly upregulated by GO exposure (**Figure 7b**). The marked GO-induced activation of the autophagic pathway was confirmed by electron and confocal microscopy analysis (**Figure 8b-d**). The number of LC3-positive vesicles in GO-treated neurons exhibited a 2-3 fold increase with respect to untreated or vehicle-treated cells and the total area covered by autophagic vesicles, quantified at CLSM level, was dramatically increased (**Figure 8e, f**).

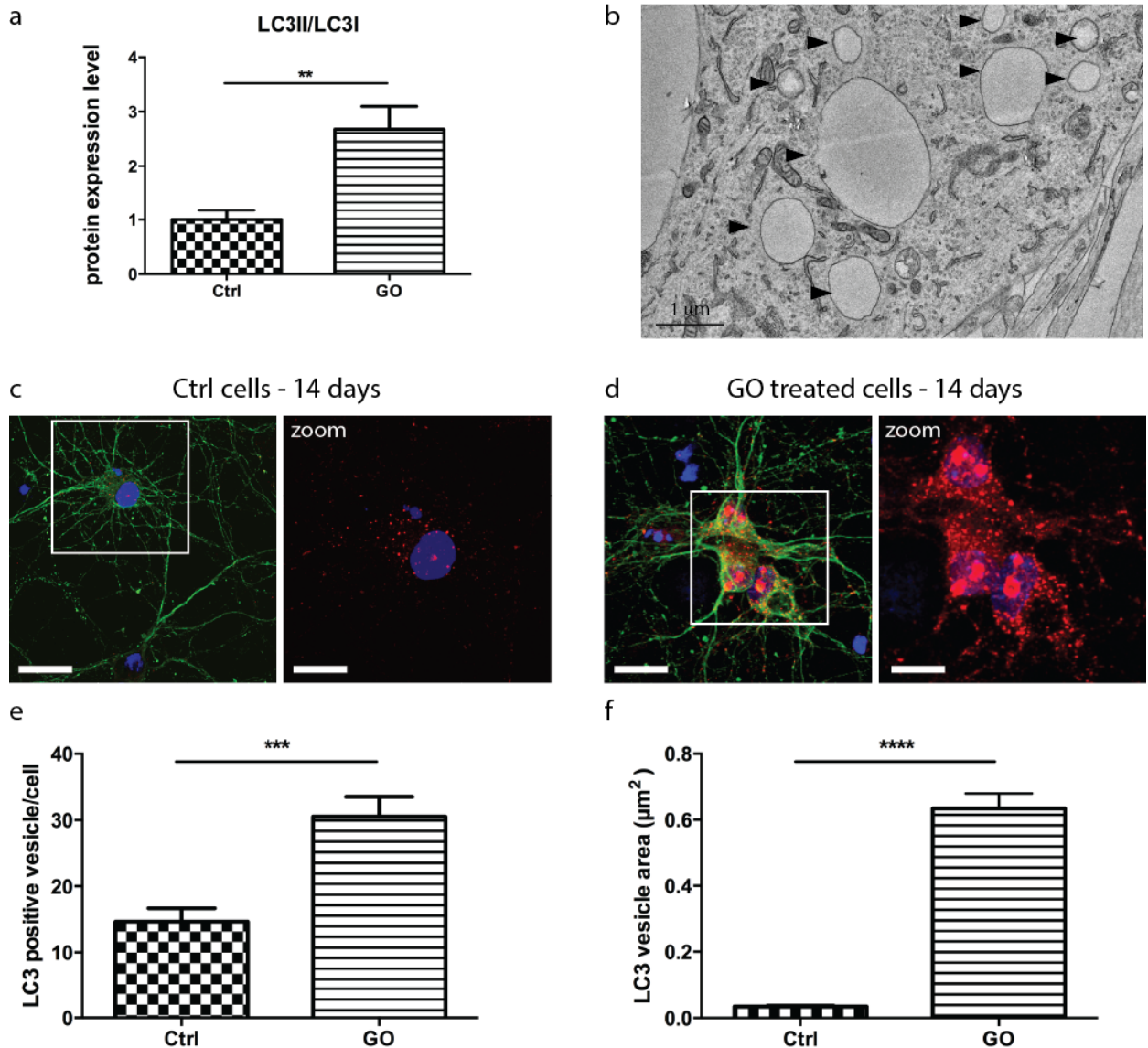

**Figure 8. Chronic exposure to GO stimulates autophagy in neurons.** (a) Quantification of Western blot analysis of MLP3A/LC3 I/II in neurons following exposure to GO flakes (means ± SEM) are shown (**$p<0.01$, two-tailed Student's *t*-test; n=3). Calnexin was used as loading control. A representative experiment is shown in **Figure 7g**. (b) Autophagy reaction upon GO exposure was also confirmed by TEM analysis (black arrowheads). (c, d) Fluorescent confocal images of neuronal cells exposed to GO flakes or vehicle for 14 days, showing nuclei in blue (DAPI), β-III tubulin staining in green and LC3-positive vesicles in red (scale bar, 20 μm; 10 μm in zoomed images). (e, f) The number of LC3-positive vesicles as well as the dimension of LC3 bodies after GO exposure compared to untreated (N.T.) or vehicle-

treated control (Ctrl) samples was quantified and found to be significantly higher in GO-treated neuronal cells (n=30 cells from 2 independent preparations; *$p<0.05$, one-way ANOVA and Bonferroni's multiple comparison test).

## DISCUSSION

In this study, we investigated the effects of the environmental exposure of primary cortical neurons to GR and GO flakes. Our analysis started from standard biocompatibility and cell morphology tests, to include a detailed characterization of neuronal functionality and flake intracellular trafficking. Moreover, proteomic and lipidomic studies allowed us to explore, with unprecedented precision, the interactions of graphene with the protein and lipid machinery of the neurons. Our aim was to achieve a comprehensive understanding of the mechanisms underlying graphene-neuron interactions, thus laying the basis for the engineering of fully biocompatible biomedical devices where graphene will directly be interfaced with neuronal cells.

One fundamental finding is that the interaction of neurons with GR/GO flakes does not affect neuronal cell viability. Neurons exposed to flakes are still able to grow and form a neuronal network after up to two weeks of GR/GO incubation with no significant changes in cell death or viability. Furthermore, also the passive and active electrophysiological properties, as well as the intrinsic excitability of neurons, were fully preserved upon both chronic and acute exposure to either GR or GO flakes, demonstrating that graphene-exposed cultures are substantially healthy. These results are in agreement with the observation that healthy neurons can be grown on planar graphene surfaces.[18]

One topic of particular interest was the physical interactions of GR/GO flakes with the neuronal membrane, which is strongly dependent on the active surface and edges of the material.[21, 28] Graphene flakes adsorbed onto the cell surface without apparently modifying or damaging the neuronal morphology. Large flake aggregates (in the μm-size order) mostly adhered to the cell membrane, while nanosheets were actively taken up and internalized by neurons. It was possible to track the trafficking of intracellular nanosheets from the early contact with the cell membrane to their final destination: graphene flakes that underwent endocytosis preferentially followed the endo-lysosomal pathway, with a significant percentage of them ending up into lysosomes after two-week incubation, in accordance with previous reports.[24] Notably GO nanosheets were more intensely endocytosed by neurons, consistent with the GO-specific changes observed by physiological, lipidomic and proteomic analyses. Although the intracellular fate of graphene flakes is important, the majority of the GR/GO flakes were not able to enter neuronal cells, presumably because of the large aggregates they formed in solution and of the relatively low ability of neurons to perform bulk endocytosis. In addition, while membrane invaginations suggestive of active uptake mechanisms were frequently observed, rare events of flakes piercing through the membrane were also noticed. The final amount of graphene that co-localized with the lysosomal marker LAMP1 was never higher than 50%, and moreover, free graphene flakes were constantly observed in the cytoplasm, leaving open the question on their final intracellular fate. Further investigations are needed to better understand the kinetics and molecular mechanisms used by graphene to enter neurons and the intracellular pathways followed by the material depending on the different sizes and surface chemistry of the graphene particles.

However, when the physiological analysis was extended to the study of network activity, intraneuronal $Ca^{2+}$ dynamics and synaptic communication, we found a clear distinct behavior of GR and GO flakes. While short- and long-term exposure to GR flakes was totally harmless for neurons and devoid of any effect on the analyzed parameters, long-term (but not short-term) contact with GO flakes had complex effects on network physiology, including: (i) overall depression of intracellular $Ca^{2+}$ dynamics, with lower resting levels, and marked inhibition of network activity as shown by the lower frequency of spontaneous and evoked $Ca^{2+}$ waves; (ii) downregulation of excitatory synaptic transmission with marked decreases in frequency and amplitude of spontaneous synaptic currents and in the density of synaptic connections; (iii) upregulation of inhibitory synapses with an increased frequency of spontaneous synaptic currents and a fully preserved complement of inhibitory synapses. The resulting imbalance between synaptic excitation and inhibition causes the depression of the network electrical activity detected by $Ca^{2+}$ imaging, an analysis that predominantly applies to excitatory neurons, which predominate under our culture conditions ($\approx 80\%$).[33] These effects are particularly relevant since GO represents the most promising graphene type for biomedical applications, because of its surface chemistry that allows high reactivity towards the biological matter and facilitates its functionalization by material chemistry.[16, 48, 49] The selective effects of GO over GR can be explained by the different chemical reactivity of the two substances, directly related to the presence of oxygen species on the surface of GO flakes, which determines their surface charge and protein binding capability, although additional effects of roughness and morphology cannot be excluded.

The functional effects of GO flakes only manifest after a long-term exposure and therefore they may result from complex neuronal responses involving transcription- and activity-dependent structural changes in neuronal networks. Indeed, a lipidomic scan revealed upregulation of PEs and downregulation of PSs. PE constitutes about 30-36% of total phospholipid content of the plasma membrane and synaptic vesicle (SV) membranes[50-52] and plays important roles in SV fusion/fission by modulating membrane curvature, synaptotagmin recruitment and formation of the SNARE fusion core complex.[53] On the other hand, PS contributes to the negative charge of the cytosolic leaflets of the membranes that regulates the fusion propensity of the plasma membrane. While the upregulation of PE could be the expression of the intense membrane trafficking and activation of the endocytic/autophagic pathways stimulated by GO, the change in the PE/PS ratio may participate in the altered SV fusion and synaptic transmission. Lipidomic analysis also emphasized a dramatic increase in the production of calcitriol/vitamin D, a neurosteroid affecting the developing and adult brain[54] and operating as a neuroprotective agent through a modulation of $Ca^{2+}$ metabolism and $Ca^{2+}$ entry through L-type $Ca^{2+}$ channels.[55] Such a dramatic upregulation of calcitriol/vitamin D is fully consistent with the depression of intracellular $Ca^{2+}$ dynamics, observed after GO exposure and particularly for the lower resting levels and accelerated decay of action potential-triggered $Ca^{2+}$ oscillations mentioned above. Consistent with these findings, the proteomic scan uncovered a strong downregulation of a series of proteins involved in $Ca^{2+}$-dependent signal transduction (CaM, CaMKII, calcineurin). Since these proteins are preferentially or exclusively (CaMKII) expressed in excitatory neurons, these changes can significantly contribute to the depression of network activity and downregulation $Ca^{2+}$ homeostasis brought about by chronic GO exposure. Interestingly, the three major interneuron-specific $Ca^{2+}$ buffers (PV, CB and CR) specifically expressed by inhibitory cortical neurons[56-58] are also markedly downregulated by GO exposure, suggesting the existence of an impaired buffering of cytosolic $Ca^{2+}$ in inhibitory neurons that can in turn facilitate inhibitory transmission.

How can the various findings be linked in a unifying pathogenic mechanism activated by GO? Chronic exposure to GO results in an imbalance between excitatory and inhibitory synaptic transmission with weakening of the former and strengthening of the latter. One possibility to explain the differential effect of GO nanosheets at excitatory and inhibitory synapses is that excitatory synapses are more vulnerable because of a higher permeability to GO flakes that can penetrate the synaptic cleft and jeopardize the exo-endocytotic cycle of SVs. However, this does not seem likely, since both excitatory and inhibitory synapses have a similar cleft thickness.[59] A more attractive possibility is based on the functional and phenotypic differences between excitatory and inhibitory neurons that can be differentially targeted by GO.

The strong impairment of excitatory transmission is characterized by a decreased frequency and amplitude of spontaneous synaptic currents and a sharp decrease in the density of excitatory synapses. The decreased mEPSC frequency parallels the reduction in synaptic density, although a decreased probability of spontaneous fusion by the decreased resting $Ca^{2+}$ levels can also be involved. The reduction in mEPSC amplitude also implies a concomitant decrease/internalization of postsynaptic AMPA-type glutamate receptors, a possible consequence of the intense endocytic processes initiated by contact with the GO flakes. This pre/post-synaptic depression of excitatory synaptic transmission could impact on the formation/maintenance of synaptic contacts, leading to adaptive pruning and reduction of the density of excitatory synapses. However, a direct effect of GO on the formation of synaptic contacts cannot be ruled out, since actin and $Ca^{2+}$-mediated processes play an important role during synaptogenesis and both of them resulted depressed by GO exposure. On the other hand, inhibitory transmission is structurally preserved and functionally enhanced by GO exposure, with an increase of mIPSCs frequency without changes in current amplitude or synapse density. Inhibitory neurons may not be affected by vitamin D overexpression and $Ca^{2+}$ downregulation. Rather, the marked downregulation of three major $Ca^{2+}$ buffers specific for inhibitory neurons, known to shape their functional properties, indicate that the increased mIPSC frequency can be ascribed to an impaired buffering of cytosolic $Ca^{2+}$ that increases the probability of spontaneous fusion. A schematic model summarizing our interpretation of the data is reported in **Figure 9**.

We envisage a stream of events that begin with alteration of lipids and $Ca^{2+}$-related proteins, which in turn affect synaptic activity resulting in an overall depressing effect on network activity. This can be initially triggered by the dysregulation of $Ca^{2+}$ homeostasis and by the intense endocytic processes initiated by contact of the plasma membrane with the GO flakes. Importantly, the observed effects are likely a combination of the cellular changes induced by internalized flakes and of the membrane modifications induced by the extracellular aggregates that are in physical contact with the plasma membrane. In this respect, the strong upregulation and activation of the autophagosome-related protein LC3, associated with a clear induction of autophagy upon GO exposure (that was recently observed in non-neuronal cells)[60, 61], suggest that neuronal networks exposed to GO tend to limit the potential damage by turning down electrical activity and activating autophagy as a stress response to the contact with graphene nanosheets.

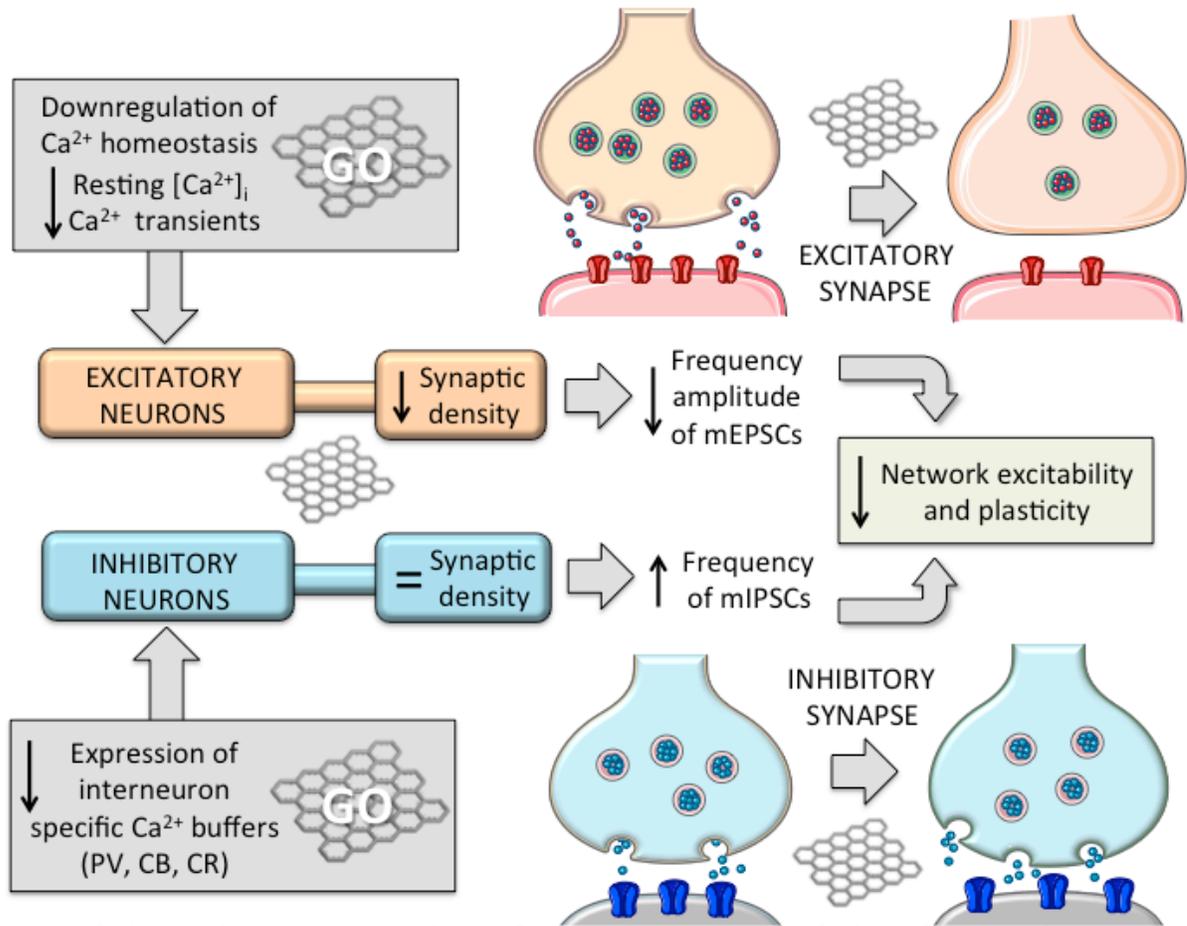

Figure 9. Schematic representation of the mechanisms of the effects of chronic exposure of GO on the activity of neuronal networks. Excitatory neurons display downregulated resting $Ca^{2+}$ levels, decreased spontaneous firing (decreased frequency of $Ca^{2+}$ transients) and decreased frequency and amplitude of spontaneous postsynaptic currents, associated with a decreased density of synaptic connections. These effects may result from calcitriol overexpression, lipidomic changes in synaptic phospholipids, increased endocytosis of postsynaptic receptors and strengthened inhibition by interneurons. On the contrary, inhibitory neurons exhibit an increased frequency of spontaneous postsynaptic currents, likely attributable to the downregulation of interneuron-specific $Ca^{2+}$-buffers, and a preserved density of synaptic connections. The resulting imbalance between excitatory and inhibitory transmission overall decreases network excitability, activity and plasticity.

## CONCLUSIONS

This work is, to our knowledge, the first to perform a comprehensive analysis of the effects of graphene flake exposure to primary neuronal cells, under controlled experimental conditions. Going beyond the mere analysis of cell viability, which has been indeed reported by several other groups,[8, 10, 15, 62] we have conducted a detailed characterization of (i) the intracellular route of internalized flakes; (ii) the effects on synaptic functionality and neuronal excitability; and (iii) the changes induced by flake exposure onto the lipid and protein machineries of the neurons. Although graphene flake exposure does not impact on cell viability and network formation, it does nevertheless have important effects on neuronal transmission and network functionality, thus warranting caution when planning to employ this material for neurobiological applications. In this respect GO, which we demonstrated to be the most biologically reactive form, is also the most largely employed graphene derivative for technological and industrial applications. It is also important to point out that although the formation of aggregates affects only to a limited extent cell survival in vitro, it may have a much higher impact in living organisms, causing tissue damage and/or inflammation reactions.[63, 64] This issue, together with our findings on the complex activity changes induced by chronic exposure to GO, will be crucial when attempting to design and engineer graphene-based devices for biomedical applications.

While the present paper was under revision, a similar paper confirming the impairment of excitatory transmission by GO was published.[65]

# METHODS

## *Synthesis and characterization of pristine graphene and graphene oxide*

Pristine graphene (GR) flakes were prepared by exfoliation of graphite through interaction with melamine by ball-milling treatment.[19] After exfoliation, melamine could be easily removed by filtration to obtain stable dispersions of few-layer graphene. In details, GR flakes were obtained by a methodology that uses mechanochemical activation by ball-milling to exfoliate graphite, through interactions with melamine (2,4,6-triamine-1,3,5-triazine) in solvent-free conditions. In a typical experiment, 7.5 mg of graphite (purchased from Bay Carbon, Inc. SP-1 graphite powder) and 0.16 mMol of melamine were ball-milled in a Retch PM100 Planetary Mill at 100 rpm for 30 min in air atmosphere. The resulting solid mixtures were dispersed in 20 ml of water to produce stable black suspensions, which were subsequently filtered and washed in hot water to remove melamine. GR/$H_2O$ dispersions were obtained at a final concentration of 0.09 mg/ml in Milli-Q-water. Melamine traces in the dispersions were estimated by Elemental Analysis (CHNS-932, Model NO: 601-800-500; LECO Corp., Saint Joseph, MI) to be 0.09 ppm. Graphene oxide (GO) was provided by Grupo Antolin Ingeniería (Burgos, Spain) by oxidation of carbon fibres (GANF Helical-Ribbon Carbon Nanofibres, GANF®) and sodium nitrate in sulfuric acid at 0 °C. While maintaining vigorous agitation, potassium permanganate was added to the suspension. After 30 min, water was slowly stirred into the paste. Then, the suspension was filtered and extensively rinsed with water to remove the presence of acids. For TEM analyses, water dispersions were placed on a copper grid (3.00 mm, 200 mesh, coated with carbon film). Samples were analyzed by High-Resolution TEM (HRTEM) using a JOEL 2100 electron microscope (JEOL, Peabody, MA). Lateral dimension distribution was calculated by using Fiji-win32. UV-vis-NIR absorbance was performed from 2.4 to 14.2 µg/ml (GR samples) or 7.5 to 20 µg/ml (GO samples) using 1 cm quartz cuvettes on a Cary 5000 UV-vis-NIR spectrophotometer (Agilent Technologies, Santa Clara, CA). For Raman spectroscopy, water dispersions were drop-cast onto a silicon surface (CZ silicon wafers; Si-Mat Silicon Materials, Kaufering, Germany). Measurements were carried out using a 100x objective at 532 nm laser excitation using a SENTERRA Raman Microscope (Bruker, Billerica, MA). An average of $I_D/I_G$ ratio was measured from different locations in the sample. The thermogravimetric analyses were performed with a TGA Q50 (TA Instruments, New Castle, DE) at 10 °C/min in a nitrogen atmosphere.

## *Preparation of primary neurons*

All experiments were carried out in accordance with the guidelines established by the European Community Council (Directive 2010/63/EU of 22 September 2010) and were approved by the Italian Ministry of Health. Primary cortical cultures were prepared from wild-type Sprague-Dawley rats (Charles River, Calco, Italy). All efforts were made to minimize suffering and to reduce the number of animals used. Rats were sacrificed by $CO_2$ inhalation, and 18-day embryos (E18) were removed immediately by cesarean section. Briefly, enzymatically dissociated cortical neurons[66] were plated on poly-D-lysine-coated (0.1 mg/ml) glass coverslips (Thermo-Fischer Scientific, Waltham, MA) at a density of 40.000 cells/ml. Cultures were incubated at 37 °C, 5% $CO_2$, 90% humidity in medium consisting of Neurobasal (Gibco/Thermo-Fischer Scientific) supplemented to reach final concentration of 5% glutamine, 5% penicillin/streptomycin and 10% B27 supplement (Gibco/Thermo-Fischer Scientific). All chemicals were purchased from Life Technologies/Thermo-Fischer Scientific unless stated otherwise. For experiments involving chronic treatments, cultures were incubated at 3 days *in vitro* (DIV) with a medium containing either 1 or 10 µg/ml of either GR or GO flakes. Controls were subjected to the same medium change with the addition of equivalent volumes of the respective vehicle (0.09 ppm melamine / $H_2O$ for GR flakes, $H_2O$ for GO flakes). Cultures were used at DIV 4, 6 and 17 (after 1, 3 and 14 days of GR / GO incubation, respectively).

## *Cell viability assay*

Primary rat cortical neurons were exposed to GR and GO (1 and 10 µg/ml), or to equivalent volumes of the respective vehicle (see above) for 24 h, 96 h and 14 days. Cells were stained with propidium iodide (PI, 1 $\mu$M) for cell death quantification, fluorescein diacetate (FDA, 2 $\mu$M) for cell viability and Hoechst 33342 (1 $\mu$M) for nuclei visualization for 3 min at room temperature (RT). Cell viability was quantified at 20x (0.5 NA) magnification using a Nikon Eclipse-80i upright epifluorescence microscope (Nikon, Tokio, Japan), with random sampling of 10 fields per sample (n = 3 coverslips/sample, from 3 independent culture preparations). Image analysis was performed using the ImageJ software and the Cell Counter plugin.

*Immunofluorescence staining*

Cortical neurons were fixed in phosphate-buffered saline (PBS) / 4% paraformaldehyde (PFA) for 20 min at RT. Cells were permeabilized with 1% Triton X-100 for 5 min, blocked with 2% fetal bovine serum (FBS) in PBS/Tween 80 0.05% for 30 min at RT and incubated with primary antibodies in the same buffer for 45 min. The primary antibodies used were: mouse monoclonal anti-β-tubulin III (#T2200, Sigma-Aldrich), mouse monoclonal anti-glial fibrillary acidic protein (GFAP, #G3893, Sigma-Aldrich), guinea pig polyclonal anti-vesicular glutamate transporter-1 (VGLUT1, #AB5905, Millipore), rabbit polyclonal anti-vesicular GABA transporter (VGAT, #131003, Synaptic System) and rabbit polyclonal anti-microtubule-associated protein 1A/1B-light chain 3 (LC3, #2775, Cell Signaling Technology). After the primary incubation and several PBS washes, neurons were incubated for 45 min with the secondary antibodies in blocking buffer solution. Fluorescently conjugated secondary antibodies were from Molecular Probes (Thermo-Fisher Scientific; Alexa Fluor 488 - #A11029, Alexa Fluor 568 - #A11036, Alexa Fluor 647 – #A21450). Samples were mounted in ProLong Gold antifade reagent with DAPI (#P36935, Thermo-Fisher Scientific) on 1.5 mm-thick coverslips.

For EEA1, LAMP1, VGLUT1/VGAT-positive terminals and LC3 vesicles, image acquisitions were performed using a confocal microscope (SP8, Leica Microsystems GmbH, Wetzlar, Germany) at 63x (1.4 NA) magnification. Z-stacks were acquired every 300 nm; 10 fields/sample (n=2 coverslips/sample, from 3 independent culture preparations). Offline analysis was performed using the ImageJ software and the JACoP plugin for co-localization studies. For each set of experiments, all images were acquired using identical exposure settings. For VGLUT and VGAT experiments, values were normalized to the relative cell volume calculated on the basis of β-tubulin III labeling.

*Scanning and Transmission Electron Microscopy*

For SEM analysis, primary cortical neurons treated with GR/GO flakes (1 and 10 μg/ml) or with the respective vehicle for 1, 3 and 14 days were fixed with 1.5% glutaraldehyde in 66 mM sodium cacodylate buffer and post-fixed in 1% $OsO_4$. Sample dehydration was performed by 5 min washes in 30%, 50%, 70%, 80%, 90%, 96% and 100% EtOH solutions. In order to fully dry the samples, overnight incubation with 99% hexamethyldisilazane (HMDS) reagent (#440191, Sigma-Aldrich) was performed. Before SEM acquisition, coverslips were sputter-coated with a 10 nm layer of 99% gold (Au) nanoparticles in an Ar-filled chamber (Cressington, Sputter Coater 208HR), and imaged using a JEOL JSM-6490LA scanning electron microscope.

For TEM analysis, primary cortical neurons treated with GR / GO flakes (1 and 10 μg/ml) or with the respective vehicle for 1, 3 and 14 days were fixed with 1.2% glutaraldehyde in 66 mM sodium cacodylate buffer, post-fixed in 1% $OsO_4$, 1.5% $K_4Fe(CN)_6$ 0.1 M sodium cacodylate, en bloc stained with 1% uranyl acetate dehydrated and flat embedded in epoxy resin (Epon 812, TAAB). After baking for 48 h at 60 °C, the glass coverslip was removed from the Epon block by thermal shock using liquid $N_2$. Neurons were identified by means of a stereomicroscope, excised from the block and mounted on a cured Epon block for sectioning using an EM UC6 ultramicrotome (Leica Microsystem, Wetzlar, Germany). Ultrathin sections (70 nm thick) were collected on copper mesh grids and observed with a JEM-1011 electron microscope operating at 100 kV and equipped with an ORIUS SC1000 CCD camera (Gatan Inc., Pleasanton, CA). For each experimental condition, at least 6 images were acquired at a magnification up to 10,000x.

*Patch-clamp Electrophysiology*

Primary rat cortical neurons exposed to GR/GO flakes (1 and 10 μg/ml) or to the respective vehicle were used for patch-clamp recordings after either 24 h or 14 days. The experiments were performed using a Multiclamp 700B amplifier (Axon Instruments, Molecular Devices, Sunnyvale, CA, USA) and an upright BX51WI microscope (Olympus, Japan) equipped with Nomarski optics. Patch electrodes fabricated from thick borosilicate glasses were pulled to a final resistance of 4-6 MΩ. Recordings with either leak current >100 pA or series resistance >20 MΩ were discarded. All recordings were acquired at 50 kHz. For the analysis of neuronal excitability, recordings in current-clamp configuration were performed in Tyrode extracellular solution in which D-(-)-2-amino-5-phosphonopentanoic acid (D-AP5; 50 $\mu$M), 6-cyano-7 nitroquinoxaline-2,3-dione (CNQX; 10 $\mu$M), Bicuculline methiodide (30 $\mu$M) and (2$S$)-3-[[(1$S$)-1-(3,4-Dichlorophenyl)ethyl] amino-2-hydroxypropyl] (phenylmethyl)phosphinic acid hydrochloride (CGP; 5 $\mu$M) were added to block NMDA, non-NMDA, $GABA_A$ and $GABA_B$ receptors, respectively. The internal solution (K-gluconate) was composed of (in mM): 126 K gluconate, 4 NaCl, 1 $MgSO_4$, 0.02 $CaCl_2$, 0.1 BAPTA, 15 Glucose, 5 Hepes, 3 ATP, 0.1 GTP, pH 7.3. Current-clamp recordings were performed at a holding potential of -70 mV, and action potential firing was induced by injecting current steps of 25 pA lasting 500 ms. All parameters were analyzed using the pClamp (Molecular Devices) and Prism6 (GraphPad Software, Inc.) softwares. Spontaneous miniature excitatory postsynaptic currents (mEPSCs) and spontaneous miniature inhibitory postsynaptic currents (mIPSCs) were recorded in voltage-clamp configuration in the presence of tetrodotoxin (TTX, 300 nM) in the extracellular solution to block the generation and propagation of spontaneous action potentials. mPSCs were acquired at 10- to 20-kHz sample frequency, filtered at half the acquisition rate with an 8-pole low-pass Bessel filter and analyzed by using the Minianalysis program (Synaptosoft, Leonia, NJ). The amplitude and frequency of mPSCs were calculated using a peak detector function using different appropriate threshold amplitude and area.

*Lipidomic and Proteomic Studies*

Solvents and chemicals were purchased from Sigma Aldrich (Milan, Italy). TMT duplex kits for peptide labeling were purchased from Thermo-Fisher Scientific. All the LC-MS instruments, columns and softwares used for metabolomics were from Waters Inc. (Milford, MA). MASCOT software for proteomics was purchased from MatrixScience Ltd. (London, UK). Pathway analysis on proteomics data was performed using publicly available REACTOME (www.reactome.org) and STRING (http://string-db.org/) software. Manual annotation of proteins for their biological functions was performed using the UNIPROT database (http://www.uniprot.org).

*Untargeted Lipidomics*

*Sample preparations.* Two independent preparations of primary cortical neurons were incubated for 2 weeks in either GO or vehicle. Eight technical replicates (0.5 million cells) per biological replicate were prepared. A reference incubation of GO without cells was also prepared and analyzed as reference blank (Medium+GO). At the end of the experiment, neurons were washed twice in serum-free medium then scraped in 0.5 ml mQ water, transferred in 2 ml tubes and immediately frozen in dry ice. Lipids were extracted by adding 1.3 ml of cold methanol (MeOH) to the samples and the tubes were vortexed for 10 min, then centrifuged for 10 min at 7000 x *g*. The supernatant solutions were then collected and the precipitates were re-extracted with 0.2 ml of MeOH-MTBE (methyl-terbutyl ether) 1:1. Supernatants from the two extractions were pooled, transferred into clean glass vials, dried under nitrogen stream and stored at -80 °C until used. On the day of analysis, samples were re-dissolved in 0.1 ml of $MeOH/CHCl_3$ solution (9:1) and transferred to glass vials for injection. Precipitates were kept and stored at -80 °C for proteomics analysis.

*Data acquisition.* Untargeted LC-MS/MS analyses were carried out on a UPLC Acquity system coupled to a Synapt G2 QToF high-resolution mass spectrometer, operating in both positive (ESI+) and negative (ESI-) ion modes. Lipids were fractionated on a reversed-phase C18 T3 column (2.1x100 mm) kept at 55 °C at a flow rate of 0.4 ml/min. The following gradient conditions were used: eluent A was 10 mM ammonium formate in 60:40 acetonitrile/water, eluent B was 10 mM ammonium formate in 90: 10 isopropyl alcohol/acetonitrile; after 1 min at 30%, solvent B was brought to 35% in 3 min, then to 50% in 1 min, and then to 100% in 13 min, followed by a 1 min 100% B isocratic step and reconditioning to 30% B. Total run time was 22 min. Injection volume was set at 4 ml. Capillary voltages were set at 3kV and 2kV for ESI+ and ESI-, respectively. Cone voltages were set at 30V for ESI+ and 35V for ESI-, respectively. Source temperature was 120 °C. Desolvation gas and cone gas ($N_2$) flows were 800 l/h and 20 l/h, respectively. Desolvation temperature was set to 400 °C. Data were acquired in MSe mode[67] with MS/MS fragmentation performed in the trap region. Low energy scans were acquired at fixed 4eV potential and high-energy scans were acquired with an energy ramp from 25 to 45eV. Scan rate was set to 0.3 s per spectrum. Scan range was set from 50 to 1200 m/z. Leucine enkephalin (2 ng/ml) was infused as lock mass for spectra recalibration.

*Data analysis.* Raw data from high-resolution LC-MS/MS runs for either polarity (ESI+ or ESI-) were subjected to a Pareto-scaled, Principal Component Analysis (PCA)[44] using the MarkerLynx software. Accurate masses (*m/z*) and retention time values (RT) were included in the multivariate analysis and assigned as X-variables (markers). The peak area for each putative marker was normalized by the total markers area and used for quantitative evaluation. Only markers observed in all the replicates were retained. Following this process, a clear group separation was observed for both polarities. The goodness of fit of the generated model was: R2= 0.839, Q2=0.809 for Principal Component 1 and R2=0.911,Q2=0.888 for Principal Component 1, clearly indicating an extremely reliable multivariate data model. The metabolic differences related to the exposure of neurons to GO were then investigated by comparing the Vehicle and the GO groups using an Orthogonal Project to Least Squares Discriminant Analysis model (OPLS-DA).[68] From the corresponding S-plot, those metabolites that showed significant up- or down-regulation between the two groups were retained and the corresponding marker list, reporting all the detected (*m/z*, RT) pairs, was generated by the software. The markers list was then exported to an Excel datasheet for further calculations, following the protocol recently described by Gonzalez-Domínguez.[69] Putative markers were then further screened by 2-tailed Student's *t*-test (Vehicle *vs* GO groups) and fold-change between the groups. A preliminary F-test was performed to discriminate equal or unequal data variances between the groups (and thus performing a homoscedastic or heteroscedastic *t*-test). Only those metabolites showing a difference p-value lower than 0.05 between the groups were retained and manually verified to eliminate spurious peaks. Retained metabolites were then identified, starting from those showing the absolute highest fold-change between the two groups. The accurate mass list was searched against the METLIN[70,71] web-based algorithm, setting an *m/z* tolerance of 5 ppm and allowing [M-H]-, [M+FA-H] - and [M-H2O-H] – for the negative ion mode and [M+H]+, [M+NH4]+, [M+Na]+, [M+K]+, and [M-H2O+H]+ for the positive ion mode as adducted species. Criteria for matching were as follows: accurate mass matching, class specific retention time and adduct type consistency, as reported by Cajka et al.[72] and, whenever possible, accurate tandem mass data matching, following the indications on MS/MS fragmentation patterns already available in the literature.[73-76]

### Untargeted Proteomics

*Sample preparation.* Protein precipitates resulting from the lipid extraction described above were used for quantitative proteomics experiments, using TMT diplex isobaric tags.[77] Briefly, following BCA protein assay,[78] 80 μg aliquots of proteins were collected from all the replicates of GO and vehicle groups, resuspended in 100 mM ammonium bicarbonate, pH 8.0, reduced with 100 mM dithiotreithol (2 ml, 60°C for 45 min), alkylated with 100 mM iodoacetamide (4 ml, room temperature for 1 h) and then digested O/N at 37 °C with proteomic-grade trypsin (1:50 in weight with the sample). Chemical labeling with TMT tags was then performed, assigning reporter 126 *m/z* to vehicle group and reporter 127 *m/z* to GO group. Labeled peptides were dried under nitrogen and resuspended in 100 ml of mobile phase A for proteomic analysis.

*Data acquisition.* Peptides were loaded on an Acquity UPLC BEH C18 (75 mm x 250 mm, 1.7 $\mu$m) nanocolumn and separated using a linear gradient of B from 3 to 45% in 3 h, followed by 10 min at 90% B and reconditioning to 3%. Flow rate was set to 350 nl/min. Total run time was 4 h. Injection volume was set at 5 $\mu$l. Eluting peptides were analyzed in positive ESI mode on a Synapt G2 QToF high-resolution mass spectrometer using a nanoLockspray ion source. The capillary voltages were set at 1.8 kV. The cone voltage was set at 27 V. The source temperature was set to 90 °C. Scan range was set from 50 to 1600 m/z. Leucine enkephalin (2 ng/ml) was infused as lock mass for spectra recalibration. Charge states 2+ to 4+ were selected for MS/MS analysis (maximum 4 precursors per MS survey scan) and dedicated collision energy profiles were automatically set by the acquisition software.

*Data analysis.* Following peak recalibration and peak-list preparation, MS/MS spectra were searched by MASCOT against the SWISSPROT database (*Rattus norvegicus* as taxonomy). TMT2plex and cysteine carbamidomethylation were selected as fixed modifications. Acetylation, methionine oxidation, phosphorylation (STY), methylation and deamidation were selected as variable modifications. A minimum of 2 unique peptides per protein and a false discovery rate below 0.05 were set for positive protein identification. At least 3 unique peptides showing an individual p-value < 0.05 were used for protein over/under-expression evaluation. Following a careful data comparison step, consisting in a trend analysis of the two biological replicates and an evaluation of the absolute differences in the up/down regulation ratios between the two replicates, the following criteria were used to generate the final list of up/down-regulated proteins: 1) hits showing opposite regulation trends between the two replicates and an absolute difference higher than 20% were discarded; 2) hits showing opposite regulation trends between the two replicates and an absolute difference lower than 20% were retained; 3) the same regulation trends were retained for any delta.

*Pathway analysis and protein annotation.* Pathway analysis on the top over (> +20%) and under (< -20%) expressed proteins was performed using the REACTOME software.[79] The calmodulin pathway analysis was performed using STRING[80] to visualize and investigate the closest functional interactors. Manual annotation of protein functions was done using the UNIPROT database.[81]

## RNA preparation and Real-time PCR analysis

Total RNA was extracted with Trizol (Invitrogen, Carlsbad, CA) according to manufacturer's instructions, purified using RNeasy MinElute Cleanup Kit (Qiagen, Hilden, Germany), and reverse transcribed into cDNA using the SuperScript III First-Strand Synthesis System (Invitrogen). Gene expression was measured by quantitative real-time PCR using CFX96 Touch Real-Time PCR Detection System (Biorad, Hercules, CA). Relative gene expression was determined using the 2-$\Delta\Delta$CT method. Gusb and Hprt1 were used as control housekeeping genes. The list of primers used is provided in **Table S4.**

## Protein Extraction and Western Blotting Analysis

Total protein lysates were obtained from cells lysed in RIPA buffer (10 mM Tris-HCl, 1 mM EDTA, 0.5 mM EGTA, 1% Triton X-100, 0.1% sodium deoxycholate, 0.1% sodium dodecyl sulfate, 140 mM NaCl) containing protease and phosphatase inhibitor cocktails (Roche, Monza, Italy). The soluble fraction was collected and protein concentration was determined using BCA Protein Assay Kit (Thermo-Fischer Scientific). For Western blotting, protein lysates were denatured at 99°C in 5X sample buffer (62.5 mM Tris-HCL, pH 6.8, 2% SDS, 25% glycerol, 0.05% bromophenol blue, 5% β-mercaptoethanol, deionized water) and separated on SDS-polyacrylamide gels (SDS-PAGE). The following antibodies were used: LC3 (0231, Nanotools, Teningen, Germany), parvalbumin (195 002, Synaptic System, Germany), calretinin (AB5054, Millipore, MA, USA), P125NAP1 (07-515; Merck, MA), calreticulin (2891, CST, Danvers, MA), CaMKII (MA1-048, Thermo-Fischer Scientific), calbindin$_{D28k}$ (300, Swant, Marly, Switzerland), calnexin (ADI-SPA 860, Enzo Life Science, Farmingdale, NY). Signal intensities

## Calcium Imaging

Cultures were loaded for 30 min at 37 °C with 1 µg/ml cell-permeable Fura-2 AM (#F1221, ThermoFisher) in the culture medium. Cells were then washed in recording buffer (10 mM HEPES pH 7.4, 150 mM NaCl, 3 mM KCl, 1 mM $MgCl_2$, 10 mM Glucose, 2 mM $CaCl_2$) for 30 min at 37 °C to allow hydrolysis of the esterified groups. Coverslips with cells were mounted on the imaging chamber and loaded with 0.5 ml recording buffer. Fura-2-loaded cultures were observed with an IX-81 motorized inverted epifluorescence microscope (Olympus, Tokyo, Japan) using a UplanSapo 60 x 1.35 NA oil-immersion objective (Olympus) and recordings were performed from visual fields containing 5 ± 2 neurons on average. During the analysis, we selected the cells by drawing regions of interest (ROI) around their bodies to reduce any background.[82] Samples were excited at 340 and 380 nm by a MT20 Hg-Xe lamp (Olympus). Excitation light was separated from the emitted light using a 395 nm dichroic mirror. Images of fluorescence emission > 510 nm were acquired continuously for a maximum of 2400 s (200 ms individual exposure time) by using a Hamamatsu Orca-ER IEEE1394 CCD camera (Hamamatsu Photonics, Hamamatsu City, Japan). The camera operated on 2x2 pixel binning mode and the imaging system was controlled by an integrating imaging software package (Cell^R; Olympus). To induce rhythmic bursts, 20 µM bicuculline methiodide was applied after 5 min recording. $Ca^{2+}$ concentration was calculated as in Grynkiewicz et al.:[83]

$$[Ca^{2+}] = Kd \times \frac{R - Rmin}{Rmax - R} \times \frac{F^{380}max}{F^{380}min}$$

where R is the measured 340/380 nm ratio; Rmin and Rmax are the ratios in absence of $Ca^{2+}$ or when Fura-2 is saturated by $Ca^{2+}$, and was determined by incubating cells in 0 $Ca^{2+}$ recording buffer with 5 mM EDTA or treating cells with 1 µM ionomycin in recording buffer containing increasing concentration of $Ca^{2+}$, ranging from 1 nM to 10 mM. $F^{380}max$ and $F^{380}min$ are the fluorescence intensity of 380 nm excitation at 0 $Ca^{2+}$ and at $Ca^{2+}$ saturation. The in situ calibration was performed by loading cells with Fura-2, then measuring the 340/380 nm ratios in $Ca^{2+}$ buffer mixtures that covered the $[Ca^{2+}]$ range of interest. To obtain the calibration curve (**Figure S11**), the values were plotted and the X-intercept representing the log of the $K_d$ in the Grynkiewicz equation was calculated by linear regression (-3,07). Recorded images were analyzed off-line by the Cell^R (Olympus, Tokio, Japan) and Prism6 softwares.

## Statistical Analysis

Data were analyzed by paired/unpaired Student's *t*-test or, in case of more than 2 experimental groups, by one-way analysis of variance (ANOVA) followed by the Bonferroni's post-hoc multiple comparison test using the Prism6 (GraphPad Software, Inc.) software. Significance level was preset to p<0.05. Data are expressed as means ± SEM throughout for number of cells (n). The normal distribution of experimental data was assessed using the Kolmogorov-Smirnov test.

Author Information
* Corresponding authors: Fabio Benfenati and Fabrizia Cesca, Center for Synaptic Neuroscience, Istituto Italiano di Tecnologia (IIT), Largo Rosanna Benzi 10, 16132 Genova, Italy; fabio.benfenati@iit.it; fabrizia.cesca@iit.it

Author Contributions
The manuscript was written through contributions of all authors. M.B. performed cell biology, electron and fluorescence microscopy and $Ca^{2+}$ imaging, immunofluorescence experiments and analysis; S.S. performed electrophysiology experiments and analysis; A.R. performed Western Blot and real-time PCR experiments; A.A. and T.B. conceived, designed and performed lipidomics and proteomics experiments and analysis; V.L and E.V. contributed to the synthesis and characterization of pristine graphene (GR) and graphene oxide (GO); F.C. and F.B. conceived the study; M.B., F.C. and F.B. conceived the experimental design and contributed to the analysis of the data; M.B., F.C., and F.B. wrote the manuscript. All authors have given approval to the final version of the manuscript.

Notes
The authors declare no competing financial interest.

Funding Sources
We acknowledge financial support from the EU FP7-ICT-2013-FET-F GRAPHENE Flagship project (no. 604391).

ACKNOWLEDGMENTS
The Electron Microscopy facility members of the Nanophysics department at the Fondazione Istituto Italiano di Tecnologia (IIT, Genova, Italy) are kindly acknowledged for use and assistance with electron imaging. We are especially grateful to Manuela Fadda and Anna Fassio for the big help in setting up the Calcium imaging experiments at the Department of Experimental Medicine of the University of Genova (Genova, Italy). Arta Mehilli is gratefully acknowledged for primary cell culture preparations, as well as Francesca Canu and Ilaria Dallorto for administrative support. The Antolin Group is also acknowledged for supporting the commercial material. We acknowledge the financial support from the EU FP7-ICT-2013-FET-F GRAPHENE Flagship project (no. 604391).

Supporting Information Available: Supplementary figures: GR and GO characterization (S1), neuronal cell viability (S2-S4), interaction of GR and GO with neurons by TEM (S5), functionality of mature neurons upon acute GR and GO exposure (S6), calcium imaging and synaptic activity in neurons upon acute GR and GO exposure (S7), synaptic activity in GR-treated neurons (S8), lipidomics data and filipin staining (S9), $Ca^{2+}$ buffering proteins in interneurons (S10); calibration curve for $Ca^{2+}$ imaging (S11); Supplementary tables: electrophysiology (1), proteomics (2-3) and RT-PCR primers (4); full "omics" data (Excel file).